\pdfoutput=1

\documentclass[ amsmath, amssymb, aps, prd, twocolumn, lengthcheck, sort&compress, showpacs, groupedaddress, nofootinbib ]{revtex4}

\usepackage{amsmath}
\usepackage{amssymb}
\usepackage{graphicx}
\usepackage{dsfont}
\usepackage{dcolumn}
\usepackage{units}
\usepackage{bm}
\usepackage{nicefrac}

\usepackage[usenames]{color}

\usepackage[colorlinks=true,linkcolor=blue,citecolor=blue,urlcolor=blue]{hyperref}

      \newcommand{\conjg}[1]{\ensuremath{\hspace{1pt}\overline{\hspace{-1pt}#1\hspace{-1pt}}}\hspace{1pt}}
      \newcommand{\vect}[1]{\bm{#1}}

\def\Slash#1{\setbox0=\hbox{$#1$} % set a box for #1
\dimen0=\wd0 % and get its size
\setbox1=\hbox{/} \dimen1=\wd1 % get size of /
\ifdim\dimen0>\dimen1 % #1 is bigger
\rlap{\hbox to \dimen0{\hfil/\hfil}} % so center / in box
#1 % and print #1
\else % / is bigger
\rlap{\hbox to \dimen1{\hfil$#1$\hfil}} % so center #1
/ % and print /
\fi}

\begin{document}

\title{Delta and Omega electromagnetic form factors\\ 
in a Dyson-Schwinger/Bethe-Salpeter approach}

\author{D.~Nicmorus}
\email{nicmorus@th.physik.uni-frankfurt.de}
\affiliation{Frankfurt Institute for Advanced Studies (FIAS), Johann Wolfgang Goethe-Universit\"{a}t, D-60438 Frankfurt am Main, Germany\\
Thomas Jefferson National Accelerator Facility, Newport News, VA 23606, USA}
\author{G.~Eichmann}
\affiliation{Institut f\"{u}r Kernphysik, Technische Universit\"at Darmstadt, D-64289 Darmstadt, Germany  }
\author{R.~Alkofer}
\affiliation{Institut f\"ur Physik, Karl-Franzens-Universit\"at Graz, A-8010 Graz, Austria }

\date{\today}

\begin{abstract}

	We investigate the electromagnetic form factors of the $\Delta$ 
	and the $\Omega$ 
	baryons within the Poincar\'{e}-covariant framework of
    Dyson-Schwinger and Bethe-Salpeter equations.
	The three-quark core contributions of the form factors are evaluated by employing a quark-diquark approximation.
   We use a consistent setup for the quark-gluon dressing, the
	quark-quark bound-state kernel and the quark-photon interaction.
	Our predictions for the multipole form factors are compatible with available experimental data and quark-model estimates.
    The current-quark mass evolution of the static electromagnetic properties agrees with results provided by lattice calculations.

\end{abstract}


\maketitle

%%%%%%%%%%%%%%%%%%%%%%%%%%%%%%%%%%%%%%%%%%%%%%%%%%%%%%%%%%%%%%%%%%%%%%%%%%%%%%%%%%%%%%%%%%%%%%%%%%%%%%%%%%%%%%%%%%%%%%%%%%%%

\section{Introduction}

        Testing the nucleon structure continues to be one of the most challenging tasks
        for contemporary experiments in particle physics. Pion, photon and electron scattering
        off nucleon targets reveal the non-pointlike nature of the nucleon
        by measuring the interactions that take place amongst the nucleon's constituents.

        The lowest-lying excited state of the nucleon, the $\Delta(1232)$ baryon, plays an equally important role.
        It is produced in such experiments at LEGS, BATES, MAMI and Jefferson Lab.
        Its mass and decay width are experimentally established. Because the $\Delta$ mainly decays into $\pi N$
        and much less into $\gamma N$, pionic effects are expected to contribute significantly to its properties.
        The very small mean lifetime of the $\Delta$ translates into a highly unstable
        electromagnetic transition $\Delta \gamma \Delta$, making a measurement of its
        electromagnetic properties very difficult.
        For example, the Particle Data Group quotes an estimate that {''\it is only a rough guess of the range''}
        for the magnetic moment $\mu_{\Delta^{++}} \simeq 3.7...7.5\,\mu_N$
        in $\pi^+ p \to \pi^+ \gamma p$ experiments~\cite{Amsler:2008zzb}.
        The MAMI result for the $\Delta^+$ magnetic moment, obtained in pion radiative photoproduction $\gamma p \to \pi^0 \gamma' p$, is
        given by $\mu_{\Delta^+} = 2.7^{+5.5}_{-5.8}\,\mu_N$ which includes both experimental and theoretical errors~\cite{Kotulla:2002cg}.
        Information on $\Delta^0$ and $\Delta^-$ static electromagnetic properties is totally missing, and
        there are no experimental results for the evolution of the charge and magnetic properties with $Q^2 \neq 0$.
        On the other hand, further insight has been achieved through the measurement of the $N \gamma \Delta$
        transition~\cite{Beck:1999ge,Pospischil:2000ad,Blanpied:2001ae,Tiator:2003xr,Sparveris:2004jn,Elsner:2005cz,Schmieden:2006xw,Stave:2006ea},
        where knowledge of the helicity amplitudes, the electric quadrupole and the Coulomb quadrupole form factors
        of the transition allows for an extraction of the $\Delta$'s electric quadrupole moment~\cite{Blanpied:2001ae}.

        Theoretically the description of the $\Delta$ has been very challenging as well.
        Its properties have been studied in quark model calculations~\cite{Isgur:1979be,Buchmann:1996bd,Faessler:2006ky,Melde:2008dg,Metsch:2003ix,Migura:2006en,Metsch:2008zz,Buchmann:2008zza,Ramalho:2008dc,Ramalho:2009vc,Ramalho:2009gk,Ramalho:2010xj},
        Skyrme models~\cite{Wirzba:1986sc,Abada:1995db,Walliser:1996ps},
        chiral cloudy bag models~\cite{Kaelbermann:1983zb,Thomas:1982kv},
        as well as chiral effective field theory~\cite{Hemmert:1996xg,Hemmert:1997ye,Pascalutsa:2004je,Hacker:2005fh,Pascalutsa:2005nd,Geng:2009ys}.
        In the absence of (accurate) experimental information, model predictions can be checked by lattice QCD.
        While the current-quark mass dependence of the baryon decuplet's static electromagnetic properties
        and related issues of their chiral extrapolation have been studied for quite some time~\cite{Leinweber:1992hy,Cloet:2003jm,Lee:2005ds,Aubin:2008qp,Boinepalli:2009sq},
        lattice results for the electromagnetic form factors' $Q^2-$evolution for a wider range of photon momenta have become available only recently~\cite{Alexandrou:2008bn,Alexandrou:2009hs,Alexandrou:2010jv}.

        Understanding the structure of the $\Delta$ baryon, its deformation from sphericity,
        and the connection to the properties of the nucleon via the $N \gamma \Delta$ quadrupole transitions
        must be complemented by extensive research of the electromagnetic
        vertex $\Delta \gamma \Delta$. Comparative studies of both
        the $N \gamma \Delta$ and $\Delta \gamma \Delta$ transition
        will reveal to which extent the deformation of the $\Delta$-baryon is provided
        by orbital angular-momentum components of its constituents.
        Naturally such a study will also shed light onto the nature of the $\Delta$ as a pure quark state,
        rather than a molecular state. Finally, perhaps the most important issue
        to be answered is the chiral cloud content of the $\Delta$-baryon.

        In connection to this, a QCD-motivated quark-core analysis of $\Delta$ electromagnetic form factors within the framework
        of Dyson-Schwinger equations (DSEs), together with hadronic bound-state equations, is 
	expected to provide further insight.
	    Dyson-Schwinger equations constitute a fully self-consistent infinite set of coupled integral equations for QCD's Green functions.
        They provide a tool to access both perturbative and non-perturbative regimes of QCD; see~\cite{Roberts:1994dr,Alkofer:2000wg,Fischer:2006ub} for reviews.
        The most prominent phenomena emerging in the latter are dynamical chiral symmetry breaking, confinement, and the formation of bound states which require
	    a nonperturbative treatment.

	    Hadrons and their properties are studied in this approach via covariant bound-state
	    equations, see~\cite{Maris:2003vk,Roberts:2007jh,Eichmann:2009zx} and references therein.
        While mesons can be described by solutions of the $q\bar{q}$-bound-state Bethe-Salpeter
     	equation (BSE), the case of a baryon is more involved.
        The three-body equivalent of the BSE is the covariant Faddeev equation.
        It was recently solved for the nucleon mass by implementing a rainbow-ladder (RL) truncation,
        i.e. a dressed gluon-ladder exchange kernel between any two quarks,
        thereby enabling a direct comparison with corresponding meson studies~\cite{Eichmann:2009qa,Eichmann:2009en}.

        In the absence of a solution for the $\Delta$ in this framework,
        a practicable simplification of the problem is based on the observation that the attractive nature of quark-antiquark
     	correlations in a color-singlet meson is also attractive for $\bar{3}_C$ quark-quark correlations
     	within a color-singlet baryon. This provides the tools for studying the three-quark problem by means of a
        covariant quark-'diquark' bound-state BSE~\cite{Ishii:1995bu,Oettel:1998bk}. At the current level of complexity, the importance of
        meson-cloud effects in the chiral and low-momentum structure of hadrons is not yet accounted for,
        hence the framework aims at a description of the hadronic quark core.

        In the present work we adopt this procedure to compute the electromagnetic properties of the $\Delta (1232)$.
        This augments our previous investigations of quark-core contributions to the $\Delta$-baryon
        mass \cite{Nicmorus:2008vb,Nicmorus:2008eh} and nucleon mass and form factors \cite{Eichmann:2008ef,Eichmann:2008kk}.
        At the same time it represents an intermediate step towards the description of the nontrivial $N\to \Delta \gamma$ transition.

        We organize the manuscript as follows: in Section~\ref{sec:qdq} we briefly summarize the Poincar\'{e}-covariant
        Faddeev approach to bary\-ons and its simplification to a quark-diquark picture; and we collect the
        ingredients of the covariant quark-diquark BSE. In Section~\ref{sec:deltaffs} we discuss the properties and construction of the $\Delta$
        electromagnetic current operator. In Section~\ref{sec:results} we 
	present and comment on the results for the $\Delta$  electromagnetic form factors and static properties.
	We also compare our results with a selection of lattice-QCD results as well as the available experimental data for the $\Delta$ and $\Omega$ baryon.
        Technical details of the calculation are collected in Appendices~\ref{app:euclidean}--\ref{app:ff-diagrams}.
        Throughout this paper we work in Euclidean momentum space and use the isospin-symmetric limit $m_u=m_d$.

%%%%%%%%%%%%%%%%%%%%%%%%%%%%%%%%%%%%%%%%%%%%%%%%%%%%%%%%%%%%%%%%%%%%%%%%%%%%%%%%%%%%%%%%%%%%%%%%%%%%%%%%%%%%%%%%%%%%%%%%%%%%

\section{Quark-diquark Faddeev-equation framework} \label{sec:qdq}

        Baryonic bound states correspond to poles in the three-quark scattering matrix.
        The three-quark bound-state amplitude is defined as the residue at the pole associated to a baryon of mass $M$.
        It satisfies a covariant homogeneous integral equation, which, upon
        neglecting irreducible three-body interactions, leads to the covariant Faddeev equation \cite{Taylor:1966zza}
        that traces the binding mechanism of three quarks in a baryon to its quark-quark correlations.

        A viable truncation of the Faddeev equation introduces diquarks as explicit degrees of freedom.
        It has been demonstrated that the same mechanism that binds color-singlet
	    mesons is suitable to account for an attraction in the corresponding diquark channels within the baryon
        \cite{Cahill:1987qr,Maris:2002yu}.
        In particular, a color-singlet baryon emerges as a bound state of a color-triplet quark and
	    color-antitriplet diquark correlations
        which are implemented via a separable sum of
	    pseudoparticle-pole contributions in the quark-quark scattering matrix.

        This procedure leads to a quark-diquark BSE on the baryon's mass shell, where
        the lightest diquarks, i.e. the scalar $0^+$ and axial-vector $1^+$ ones, have been used to describe
	    the nucleon. The spin$-\nicefrac{3}{2}$ and isospin$-\nicefrac{3}{2}$ flavor symmetric $\Delta$ necessitates only
        axial-vector diquark correlations; its quark-diquark BSE reads \cite{Oettel:1998bk,Oettel:2000jj}
                   \begin{equation}\label{deltabse}
                        \Phi^{\mu\nu}(p,P) = \!\!\int\limits_k
			            K^{\mu\rho}(p,k,P) \, S(k_q) \, D^{\rho\sigma}(k_d) \,\Phi^{\sigma\nu}(k,P)   \,,
                   \end{equation}
	    where $P$ is the total baryon momentum, $k_q$, $k_d$ are quark and diquark momenta, $p$, $k$ are the
	    quark-diquark relative momenta, and $\int_k$ denotes $\int d^4 k/(2\pi)^4$.
        Greek superscripts represent Lorentz indices, Greek subscripts fermion indices.
        The amplitudes $\Phi^{\mu\nu}_{\alpha\beta}(p,P)$ are the matrix-valued remainders of the full
        quark-diquark amplitude $\Phi_{\alpha\beta}^{\mu\nu}(p,P)\,u^\nu_\beta(P)$ for the $\Delta$, where $u^\nu_\beta(P)$ is a Rarita-Schwinger spinor
        describing a free spin-3/2 particle with momentum $P$.

        \begin{figure}[tp]
                    \begin{center}

                    \includegraphics[scale=1.7]{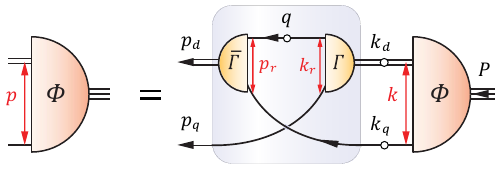}
                    \caption{(Color online) The quark-diquark BSE, Eq.\,(\ref{deltabse})}\label{fig1}

                    \end{center}
        \end{figure}

         \begin{figure*}[htp]
                    \begin{center}

                    \includegraphics[scale=1.31]{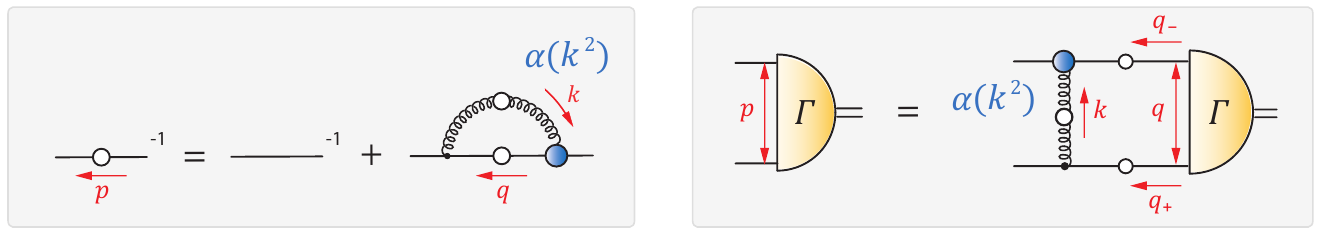}
                    \caption{(Color online) Quark DSE \eqref{quarkdse} and diquark BSE \eqref{dqbse} in rainbow-ladder truncation.}\label{fig:dsebse}

                    \end{center}
        \end{figure*}

        In order to solve Eq.~(\ref{deltabse}) one needs to specify the dressed-quark propagator $S$, the axial-vector diquark propagator $D^{\rho\sigma}$,
	    and the axial-vector diquark amplitude $\Gamma^\nu$ and its charge-conjugate $\conjg{\Gamma}^\nu$ which appear
	    in the quark-diquark kernel:
            \begin{equation}\label{deltabse-kernel}
                 K^{\mu\nu}(p,k,P) =  \Gamma^\nu(k_r,k_d) \, S^T(q) \, \conjg{\Gamma}^\mu(p_r,p_d) \,,
            \end{equation}
        where subscripts $"r"$ denote quark-quark relative momenta and $"d"$ diquark momenta.
        The mechanism which binds the $\Delta$ and is expressed through Eqs.~(\ref{deltabse}--\ref{deltabse-kernel}) is
	    an iterated exchange of roles between the single quark and any of the quarks contained in the diquark.
        This exchange is depicted in Fig.~\ref{fig1}.

        A solution of the quark-diquark BSE for the $\Delta$ was presented in Ref.~\cite{Nicmorus:2008vb} and corresponding results for its mass were reported therein.
        In the following subsections we proceed by recollecting the ingredients of Eq.\,\eqref{deltabse}.

%%%%%%%%%%%%%%%%%%%%%%%%%%%%%%%%%%%%%%%%%%%%%%%%%%%%%%%%%%%%%%%%%%%%%%%%%%%%%%%%%%%%%%%%%%%%%%%%%%%%%%%%%%%%%%%%%%%%%%%%%%%%

\subsection{Quark propagator and quark-gluon coupling} \label{sec:qprop}

    The fundamental building block which appears in Eqs.\,(\ref{deltabse}--\ref{deltabse-kernel}) and connects the quark-diquark model and resulting hadron properties
    with the underlying structure of QCD is the dressed quark propagator $S(p)$.
    It is expressed in terms of two scalar functions,
            \begin{equation}\label{qprop}
                S^{-1}(p) = A(p^2)\,\left( i\Slash{p} + M(p^2) \right),
            \end{equation}
    namely the quark wave-function renormalization $1/A(p^2)$ and the quark mass function $M(p^2)$.
    Dynamical chiral symmetry breaking becomes manifest through a non-perturbative enhancement
    of both dressing functions $M(p^2)$ and $A(p^2)$ at small momenta which indicates
    the dynamical generation of a large constituent-quark mass.

    Such a dynamical enhancement emerges in the solution of the quark DSE, cf. Fig.\,\ref{fig:dsebse}:
            \begin{equation}\label{quarkdse}
                S^{-1}_{\alpha\beta}(p) = Z_2 \left( i\Slash{p} + m \right)_{\alpha\beta}  +
		               \int\limits_q \mathcal{K}_{\alpha\alpha'\beta'\beta}(p,q) \,S_{\alpha'\beta'}(q)\,,
            \end{equation}
    where $Z_{2}$ is the quark renormalization constant and $m$ the bare current-quark mass which
    constitutes an input of the equation.
	The interaction kernel $\mathcal{K}$ includes the dressed gluon propagator as well as one bare
	and one dressed quark-gluon vertex.

    In principle, the dressed gluon propagator and quark-gluon vertex could
	be obtained as solutions of the infinite coupled tower of QCD's DSEs, together with all
	other Green functions of the theory.
	In practical numerical studies one employs a truncation: only a subset of the
	infinite system of equations is solved for explicitly; Green functions appearing in the
	subset but not solved for are represented by substantiated ans\"atze.

    In connection with meson properties, e.g. to establish the pion as the Goldstone boson of spontaneous chiral symmetry breaking,
    it is imperative to employ a truncation that preserves the axial-vector Ward-Takahashi identity.
    The latter connects the kernel of the quark DSE with that of a meson BSE,
    ensures a massless pion in the chiral limit and leads to a generalized Gell-Mann--Oakes--Renner relation
    \cite{Maris:1997hd,Holl:2004fr}.
        Such a symmetry-preserving truncation scheme was described in~\cite{Munczek:1994zz,Bender:1996bb},
        and its lowest order is
        the rainbow-ladder (RL) truncation which amounts to an iterated dressed-gluon exchange between quark and antiquark.
        It has been extensively used in Dyson-Schwinger studies of hadrons, see e.g.~\cite{Krassnigg:2009zh,Eichmann:2007nn} and references therein.
    The RL truncation retains only the vector part $\sim \gamma^\mu$ of the dressed quark-gluon vertex.
    Its non-perturbative dressing, together with that of the gluon propagator, is absorbed into an effective coupling $\alpha(k^2)$ which is modeled.
	The kernel $\mathcal{K}$ of both quark DSE and meson BSE then reads:
            \begin{equation}\label{RLkernel}
                \mathcal{K}_{\alpha\alpha'\beta\beta'} =  Z_2^2 \, \frac{ 4\pi \alpha(k^2)}{k^2} \, T^{\mu\nu}_k \gamma^\mu_{\alpha\alpha'} \,\gamma^\nu_{\beta\beta'},
            \end{equation}
    where $T^{\mu\nu}_k=\delta^{\mu\nu} - \hat{k}^\mu \hat{k}^\nu$
    is a transverse projector with respect to the gluon momentum $k=q-p$, and $\hat{k}^\mu=k^\mu/\sqrt{k^2}$
	denotes a normalized 4-vector.

\begin{figure}[t]
                    \begin{center}

                    \includegraphics[scale=1.25]{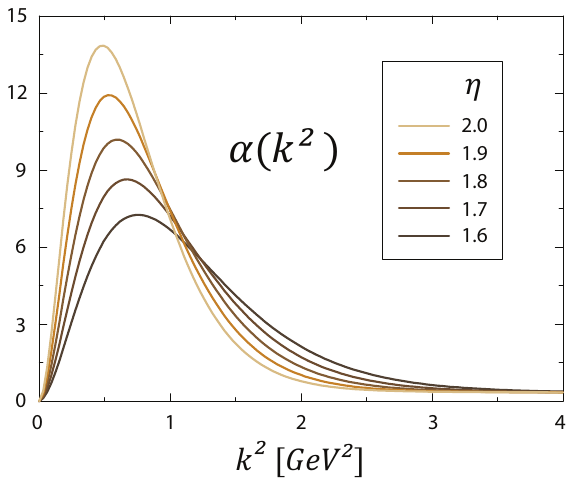}
                    \caption{(Color online) Effective coupling $\alpha(k^2)$ of Eq.\,\eqref{couplingMT}, evaluated for $\Lambda = 0.98$ GeV
                                            corresponding to the $u/d-$quark mass and in the range $\eta \in [1.6,2.0]$.}\label{fig:coupling}

                    \end{center}
        \end{figure}

    At large gluon momenta, the effective coupling $\alpha(k^2)$ is constrained by perturbative QCD;
    in the deep infrared, its behavior is irrelevant for hadronic ground states~\cite{Blank:2010pa}.
    At small and intermediate momenta it must exhibit sufficient strength to
    allow for dynamical chiral symmetry breaking and the dynamical
	generation of a constituent-quark mass scale.
    We employ the frequently used ansatz~\cite{Maris:1999nt}
                    \begin{equation}\label{couplingMT}
                        \alpha(k^2) = \pi \eta^7  \left(\frac{k^2}{\Lambda^2}\right)^2 \!\! e^{-\eta^2 \left(\frac{k^2}{\Lambda^2}\right)} + \alpha_\text{UV}(k^2) \,,
                    \end{equation}
        where the second term reproduces the logarithmic decrease of QCD's perturbative running coupling and vanishes at $k^2=0$.
        The first term supplies the necessary infrared strength and is characterized by two parameters: an infrared scale $\Lambda$ and a dimensionless width parameter $\eta$, cf. Fig.~\ref{fig:coupling}.
        (They are related to the infrared parameters of Ref.~\cite{Nicmorus:2008vb} via
        $c= (\Lambda/\Lambda_0)^3$ and $\omega = \eta^{-1} \Lambda/\Lambda_0$, with $\Lambda_0 = 1$~GeV.)

        The interaction of Eq.~(\ref{couplingMT}) provides a reasonable description of pseudoscalar-meson, vector-meson and
        nucleon ground-state properties if the scale $\Lambda$ is adjusted to reproduce the experimental pion decay constant
        and kept fixed for all values of the quark mass (see~\cite{Maris:2006ea,Krassnigg:2009zh,Nicmorus:2008vb} and references therein). The corresponding value is $\Lambda = 0.72$~GeV.
        Furthermore, these observables have turned out to be insensitive to the shape of the coupling in the infrared \cite{Maris:1999nt,Krassnigg:2009zh}; i.e.,
        to a variation of the parameter $\eta$ around the value $\eta \approx 1.8$.

        Other quantities, most notably the masses of axial-vector and pseudoscalar isosinglet mesons, are not reproduced so well in a RL truncation.
        Efforts to go beyond RL have been made, and are underway (see e.g.~\cite{Alkofer:2008tt,Fischer:2009jm,Chang:2009zb}), but typically require a significant amplification of numerical effort.
        In recent studies certain additional structures in the quark-gluon vertex, and for consistency also in the quark-antiquark kernel, have proven capable
        to provide a better description of such observables as well~\cite{Alkofer:2008et,Fischer:2008wy,Fischer:2009jm,Chang:2009zb,Chang:2010jq}.

        On the other hand, substantial attractive contributions come from a pseudoscalar meson cloud which augments the 'quark core' of dynamically
	    generated hadron observables in the chiral regime, whereas it vanishes with increasing current-quark mass.
        A viewpoint explored in Ref.~\cite{Eichmann:2008ae} was to identify RL with the quark core of chiral effective field theory
        which, among other corrections, must be subsequently dressed by pion-cloud effects.
        From this perspective a coincidence of RL results in the chiral region with experimental or lattice data becomes objectionable.
        The properties of a hadronic quark core were then mimicked by implementing a current-mass dependent scale $\Lambda(m)$ which is deliberately inflated close to the chiral limit,
        where $\Lambda \approx 1$ GeV.
         As a result, mass-dimensionful $\pi$, $\rho$, $N$ and $\Delta$ observables were shown to be consistently overestimated and mostly
         compatible with quark-core estimates from quark models and chiral perturbation theory;
         for a detailed discussion, see \cite{Eichmann:2008ae,Eichmann:2008ef,Nicmorus:2008vb}.

        In the present work we employ this 'core model' of Ref.~\cite{Eichmann:2008ae} to compute the $\Delta\gamma\Delta$ transition properties.
        However, as we will argue in Section~\ref{sec:results}, the distinction between the core model ($\Lambda(m)$) and the fixed-scale version ($\Lambda = 0.72$ GeV) becomes mostly irrelevant
        once the scale of dynamical chiral symmetry breaking (here: the mass of the $\Delta$) has been set and all dimensionful quantities are expressed in terms of this scale.
        We finally stress that through Eq.\,\eqref{couplingMT} all parameters of the interaction $\alpha(k^2)$ are fixed by using information from $\pi-$ and $\rho-$meson core properties only.

%%%%%%%%%%%%%%%%%%%%%%%%%%%%%%%%%%%%%%%%%%%%%%%%%%%%%%%%%%%%%%%%%%%%%%%%%%%%%%%%%%%%%%%%%%%%%%%%%%%%%%%%%%%%%%%%%%%%%%%%%%%%

\subsection{Diquarks}

        Various theoretical approaches as well as experimental observations indicate
        that the strong attraction between two quarks to form diquarks \textit{within} a baryon
        is a key feature for a better understanding of hadron properties~\cite{Anselmino:1992vg,Jaffe:2004ph}.
        This entails that the quark-quark scattering matrix is dominated by diquark degrees of freedom
        at small spacelike and timelike values of the total two-quark momentum $P$.
        A certain singularity structure in the timelike region indicates the presence of diquark mass scales within a baryon.
        In the simplest case such a structure can be realized through timelike diquark poles at certain values of $P^2$, i.e.
        $P^2=-m_\text{sc}^2$, $P^2=-m_\text{av}^2$, which characterize the lightest diquarks, namely the scalar and axial-vector ones.
        Diquarks carry color and are hence not observable; yet such a pole structure does per se not contradict diquark confinement, see e.g.~\cite{Alkofer:2000wg}.

        In the present context, timelike diquark poles emerge as an artifact of the RL truncation which does not persist beyond RL~\cite{Bender:1996bb}.
        In complete analogy to a meson or baryon case, it nevertheless allows to derive bound-state equations at these poles which
        determine the on-shell scalar and axial-vector diquark amplitudes that appear in Eq.\,\eqref{deltabse-kernel}, together with their masses.
        The assumption that this separable structure of the scattering matrix persists for \textit{all} values of $P^2$ is the underlying condition
        which simplifies the Faddeev equation to a quark-diquark model.

        Compared to the nucleon, only an isospin-$1$ diquark can contribute to the isospin-$\nicefrac{3}{2}$ $\Delta$ amplitude
        which excludes the involvement of a scalar diquark.
        The Bethe-Salpeter equation for the on-shell axial-vector diquark amplitude $\Gamma^\mu$ reads
                    \begin{equation}\label{dqbse}
                        \Gamma^\mu_{\alpha\beta}(p,P) =\int\limits_q
                        \mathcal{K}_{\alpha\alpha'\beta\beta'} \left\{ S(q_+) \Gamma^\mu(q,P) S^T(q_-) \right\}_{\alpha'\beta'},
                    \end{equation}
        where $P$ is the diquark momentum, $p$ is the relative momentum between the two quarks in the diquark bound state
        and $q_\pm = \pm q + P/2$ are the quark momenta.
        The equation has the same shape as a vector-meson BSE; in a RL truncation the kernel $\mathcal{K}$ is given by Eq.\,\eqref{RLkernel}.
        The inherent color structure of the kernel leads to prefactors $4/3$ and $-2/3$ for the integrals in \eqref{quarkdse} and \eqref{dqbse}, respectively.
        Poincar\'e covariance entails that the axial-vector diquark amplitude does not only consist of its dominant structure $\sim\gamma^\mu C$,
        but involves 12 momentum-dependent basis elements which are self-consistently generated upon solving Eq.\,\eqref{dqbse}.

        The diquark BSE only specifies the on-shell diquark amplitude, i.e. at $P^2=-m_\text{av}^2$.
        Diquarks in a baryon are offshell. Information on the off-shell behavior of the scattering matrix $T$
        can be inferred from its Dyson series, schematically written as $T = \mathcal{K} + \int \mathcal{K} SS \,T$. Reinserting
        the separable ansatz $T  = \Gamma^\mu  D^{\mu\nu}  \Gamma^\nu $ yields an expression for the axial-vector diquark propagator $D^{\mu\nu}(P)$:
            \begin{equation} \label{dqprop}
                D^{-1}_{\mu\nu}(P) = m_\text{av}^2 \left\{ \lambda \,\delta_{\mu\nu} +
                                   \beta\,F_{\mu\nu}(P) + Q_{\mu\nu}(P) \right\},
            \end{equation}
        where $Q_{\mu\nu}$ and $F_{\mu\nu}$ are one- and two-loop integrals involving diquark amplitudes, the quark propagator
        and the RL kernel $\mathcal{K}$, and $\lambda$ and $\beta$ are related to their on-shell values; see \cite{Nicmorus:2008vb,Eichmann:2009zx} for details.
        Eq.\,\eqref{dqprop} completely specifies the diquark propagator from its substructure.
        On the mass shell, it behaves like a transverse particle pole: $D^{-1}_{\mu\nu}(P^2=-m_\text{av}^2) = (P^2 + m_\text{av}^2)\, T^{\mu\nu}_P$,
        whereas for offshell momenta, via Dyson's equation, it picks up non-resonant contributions which are implicit in the T-matrix but cannot be described by a free-particle propagator.

%%%%%%%%%%%%%%%%%%%%%%%%%%%%%%%%%%%%%%%%%%%%%%%%%%%%%%%%%%%%%%%%%%%%%%%%%%%%%%%%%%%%%%%%%%%%%%%%%%%%%%%%%%%%%%%%%%%%%%%%%%%%

         \begin{figure}[t]
                    \begin{center}

                    \includegraphics[scale=1.38]{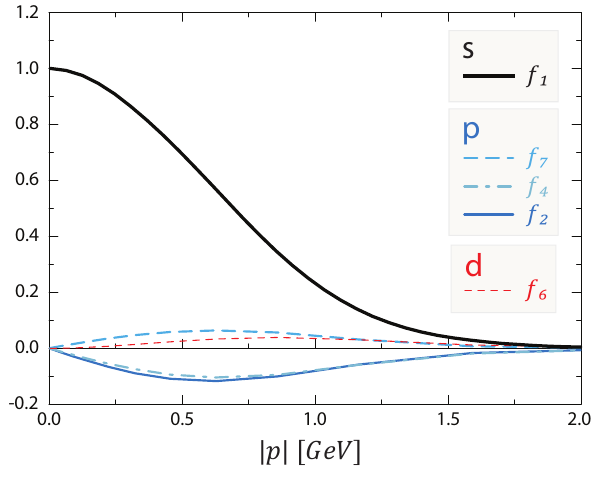}
                    \caption{(Color online) Result for the quark-diquark $\Delta$ amplitude.
                                            The plot shows the zeroth Chebyshev moments of the dressing functions
                                            $f_k^\Delta$ in Eq.\,\eqref{deltaamplitdecompos}, where the labeling
                                            corresponds to Eq.\,18 of Ref.~\cite{Nicmorus:2008vb}.
                                            The remaining components $f_3$, $f_5$ and $f_8$ are small and not displayed.}\label{fig:amplitude}

                    \end{center}
        \end{figure}

\subsection{Quark-diquark $\Delta$ amplitudes}

        All ingredients of the quark-diquark BSE \eqref{deltabse} are now specified:
        the quark propagator as obtained from the quark DSE \eqref{quarkdse}, the diquark amplitude as a solution of the diquark BSE \eqref{dqbse},
        and the diquark propagator from Eq.\,\eqref{dqprop}; all obtained within a RL truncation that involves the effective coupling $\alpha(k^2)$.
        To compute the amplitude and mass of the $\Delta$ baryon numerically,
        the structure of the on-shell quark-diquark amplitude $\Phi^{\mu\nu}$ must be specified.
        It is decomposed into 8 covariant and orthogonal basis elements:
                    \begin{equation}\label{deltaamplitdecompos}
                        \Phi^{\mu\nu}(p,P) = \sum_{k=1}^8 f_k^\Delta(p^2,\hat{p}\cdot\hat{P}) \,\tau_k^{\mu \rho}(p,P)\,\mathds{P}^{\rho \nu}(P)\,,
                    \end{equation}
        where $P$ is the $\Delta$ onshell momentum with $P^2=-M_\Delta^2$ and $\hat{P}= P/(iM_\Delta)$.
        The Rarita-Schwinger projector onto positive energy and spin-$3/2$ is given by
                    \begin{equation}\label{RSprojector}
                        \mathds{P}^{\rho \nu}(P) = \Lambda_+(P)\left( T^{\rho \nu}_P - \frac{1}{3}\,\gamma_T^\rho\,\gamma_T^\nu\right),
                     \end{equation}
        where $\Lambda_+(P) =(\mathds{1} + \hat{\Slash{P}})/2$ is the positive-energy projector,
        $T^{\alpha\beta}_P = \delta^{\alpha\beta} - \hat{P}^\alpha \hat{P}^\beta$ is a transverse projector with respect to the total momentum, and
        $\gamma_T^\alpha = T^{\alpha\beta}_P \gamma^\beta$ are transverse $\gamma-$matrices.

        The details of the calculation as well as results for the $\Delta$ mass were reported in Ref.~\cite{Nicmorus:2008vb}.
        A partial-wave analysis of the quark-diquark amplitude assigns total quark-diquark spin and orbital angular momentum quantum numbers to each of the 8 basis elements.
        Upon performing the flavor-color traces, a standard procedure to solve the quark-diquark BSE
        involves a Chebyshev expansion in the angular variable $\hat{p} \cdot \hat{P}$ and
        leads to coupled one-dimensional eigenvalue equations for the Chebyshev moments of the dressing functions $f_k^\Delta$.
        They match the BSE solution at $P^2=-M_\Delta^2$, i.\,e.~for an eigenvalue $\lambda_\text{BSE}(P^2=-M_\Delta^2)=1$.

        The $\Delta$ mass in Ref.~\cite{Nicmorus:2008vb} was calculated for both model versions discussed in Sec.~\ref{sec:qprop}.
        Using a fixed scale $\Lambda=0.72$ GeV yields the result $M_\Delta=1.28$ GeV which is reasonably close to the experimental value
        $1.232$ GeV.
        In the core version, where $\Lambda=0.98$ GeV at the $u/d$ mass, the result is $M_\Delta=1.73(5)$ GeV, where the bracket denotes the sensitivity to the
        infrared width parameter $\eta$.
        From the perspective of chiral effective field theory, pionic effects
        should reduce the $\Delta$ mass by merely $\sim 300$ MeV which might indicate the relevance of further diquark channels in describing $\Delta$ properties.
        On the other hand the 'core' $\Delta$ in the present approach is not a resonance since the $\Delta\to N\pi$
        decay channel is not accounted for in the quark-diquark kernel, and a corresponding non-zero width might impact on its mass as well.

       The result for the $\Delta$ amplitude, as obtained in the RL-truncated quark-diquark approach,
       is dominated by an $s-$wave component in its rest frame, cf. Fig.~\ref{fig:amplitude}.
       The subleading $s-$, $p-$ and $d-$wave amplitude components are significantly suppressed
       compared to this structure which corresponds to $\tau_1^{\mu\rho} = \delta^{\mu\rho}$ in Eq.\,\eqref{deltaamplitdecompos}.
       A similar observation holds for the nucleon amplitude and
       might indicate that orbital angular-momentum correlations in these baryons' amplitudes are dominated
       by pionic effects which are absent in our setup.

%%%%%%%%%%%%%%%%%%%%%%%%%%%%%%%%%%%%%%%%%%%%%%%%%%%%%%%%%%%%%%%%%%%%%%%%%%%%%%%%%%%%%%%%%%%%%%%%%%%%%%%%%%%%%%%%%%%%%%%%%%%%

         \begin{figure*}[htp]
                    \begin{center}
                    \includegraphics[scale=0.93]{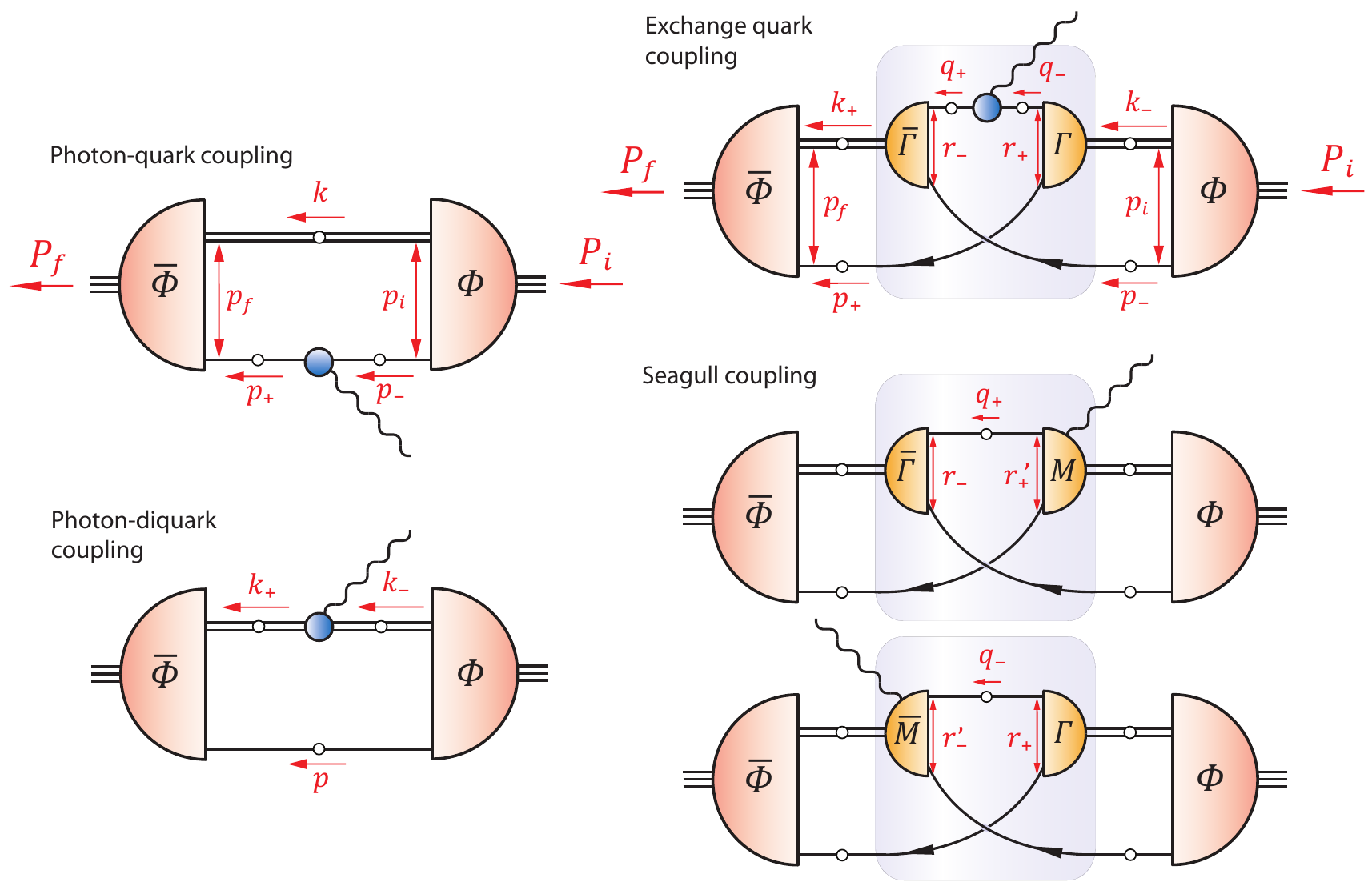}
                    \caption{(Color online) A baryon's electromagnetic current in the quark-diquark model.}\label{fig2}
                    \end{center}
        \end{figure*}

\section{Delta electromagnetic form factors} \label{sec:deltaffs}

\subsection{Electromagnetic current operator}

 \renewcommand{\arraystretch}{1.2}

             Having numerically calculated the $\Delta$-baryon amplitudes, we proceed with
             the construction of the $\Delta$ electromagnetic current.
             It can be written in the form
             \begin{align}\label{eq:current2}
                 &J^{\mu,\rho\sigma}(P,Q) = i\,\mathds{P}^{\rho\alpha}(P_f) \left[
                          \left( F_1^\star\,\gamma^\mu - F_2^\star\,\frac{\sigma^{\mu\nu}Q^\nu}{2M_\Delta}\right)\delta^{\alpha\beta}  \right.   \nonumber \\
                           &\left. -\left( F_3^\star\,\gamma^\mu - F_4^\star\,\frac{\sigma^{\mu\nu}Q^\nu}{2M_\Delta} \right) \frac{Q^\alpha Q^\beta}{4M_\Delta^2}
                           \right]\mathds{P}^{\beta\sigma}(P_i)
             \end{align}
             which is derived in App.~\ref{app:currentderivation}.
             The exchanged photon momentum is denoted by $Q=P_f-P_i$, where $P_i$ and $P_f$ are the initial and final momenta of the $\Delta$
             and $P=(P_i+P_f)/2$ is its average total momentum.
             The Rarita-Schwinger projectors were defined in Eq.\,\eqref{RSprojector}.

             The electromagnetic current is expressed in terms of four form factors $F_i^\star(Q^2)$.
             The experimentally measured $\Delta$ form factors -- Coulomb monopole $G_{E0}$, magnetic dipole $G_{M1}$,
             electric quadrupole $G_{E2}$, and magnetic octupole $G_{M3}$ -- can be expressed through linear combinations of the $F_i^\star(Q^2)$~\cite{Nozawa:1990gt,Pascalutsa:2006up}:

             \begin{align}
               G_{E_0} &:= \left(1+\frac{2\tau}{3}\right) ( F_1^\star - \tau F_2^\star)  - \frac{\tau}{3} (1+\tau) \,( F_3^\star - \tau F_4^\star) \,, \nonumber \\
               G_{M_1} &:= \left(1+\frac{4\tau}{5}\right) (F_1^\star+F_2^\star) - \frac{2\tau}{5} (1+\tau)\, (F_3^\star + F_4^\star)\,, \nonumber \\
               G_{E_2} &:= (F_1^\star - \tau F_2^\star) - \frac{1}{2}\,(1+\tau) \,(F_3^\star - \tau F_4^\star)\,, \nonumber \\
               G_{M_3} &:= (F_1^\star + F_2^\star) - \frac{1}{2}\,(1+\tau) \,(F_3^\star + F_4^\star)\,.
             \end{align}
             Their static dimensionless values are given by
             \begin{equation}
                 \begin{array}{l} G_{E_0}(0) = e_\Delta\,, \\ G_{E_2}(0) = \mathcal{Q}\,, \end{array}\qquad
                 \begin{array}{l} G_{M_1}(0) = \mu_\Delta \,, \\ G_{M_3}(0) = \mathcal{O}\,, \end{array}
             \end{equation}
             where $e_\Delta \in \{  2,1,0,-1 \}$ is the $\Delta$ charge, $\mu_\Delta$ its magnetic dipole moment,
             $\mathcal{Q}$ the electric quadrupole moment, and $\mathcal{O}$ the magnetic octupole moment. Equivalently, one has
             \begin{equation}
                 \begin{array}{l} F_1^\star(0) = e_\Delta\,, \\ F_3^\star(0) = e_\Delta -\mathcal{Q}\,, \end{array}\quad
                 \begin{array}{l} F_2^\star(0) = \mu_\Delta-e_\Delta\,, \\ F_4^\star(0) = \mu_\Delta-e_\Delta + \mathcal{Q} - \mathcal{O}\,. \end{array}
             \end{equation}
             These form factors are dimensionless. Their dimensionful values are given by

             \begin{equation*}
                 G_{E_2}^\text{dim} = \frac{e\,G_{E_2}}{M_\Delta^2}\,, \quad
                 G_{M_1}^\text{dim} = \frac{e\,G_{M_1}}{2M_\Delta}\,, \quad
                 G_{M_3}^\text{dim} = \frac{e\,G_{M_3}}{2M_\Delta^3}\,.
             \end{equation*}

%%%%%%%%%%%%%%%%%%%%%%%%%%%%%%%%%%%%%%%%%%%%%%%%%%%%%%%%%%%%%%%%%%%%%%%%%%%%%%%%%%%%%%%%%%%%%%%%%%%%%%%%%%%%%%%%%%%%%%%%%%%%%

       \begin{figure*}[tp]
                    \begin{center}

                    \includegraphics[scale=1.68]{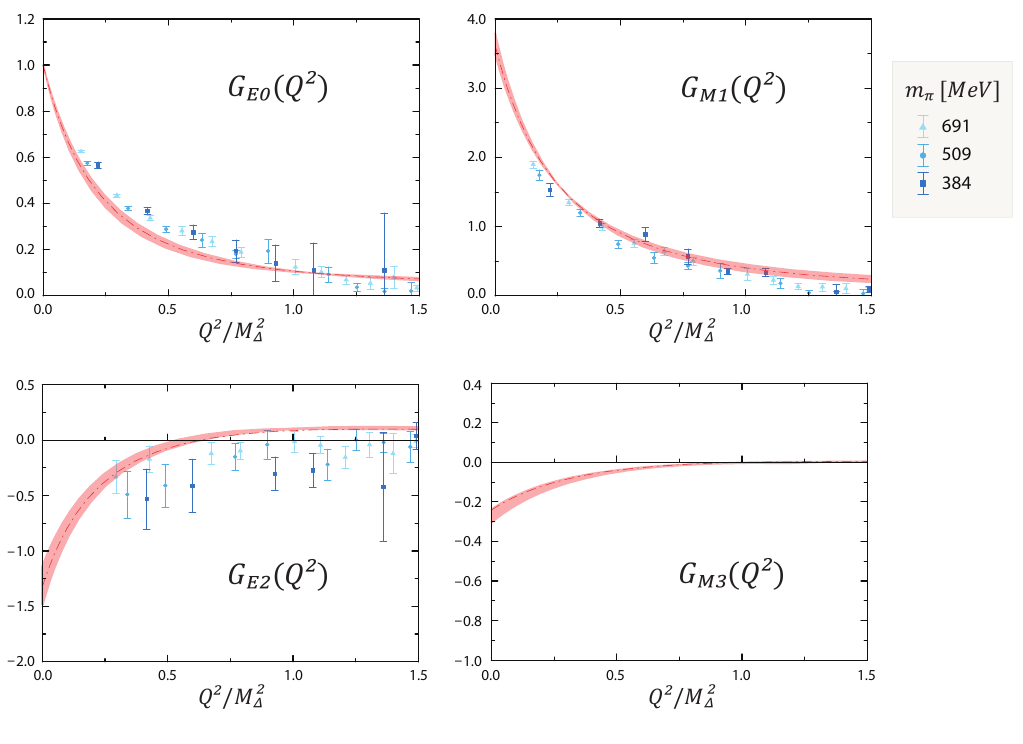}
                    \caption{(Color online) Electromagnetic form factors of the $\Delta$.
                    The bands represent the sensitivity to a variation of $\eta=1.8\pm 0.2$.
                    The results are compared to unquenched lattice data of Ref.~\cite{Alexandrou:2009hs} at three different pion masses.} \label{fig3}

                    \end{center}
        \end{figure*}

\subsection{Construction of the electromagnetic current}

             To compute the electromagnetic properties of the $\Delta$-baryon in a given framework, one must specify how the photon
             couples to its constituents.
             In the quark-diquark context this amounts to resolving the coupling of the photon to the dressed quark, to the diquark, and to the interaction between them,
             where the incoming and outgoing baryon states are described by the quark-diquark amplitudes of Eq.\,\eqref{deltaamplitdecompos}.

             The construction of this current is based on a procedure which automatically satisfies electromagnetic gauge invariance~\cite{Kvinikhidze:1999xp,Oettel:1999gc}.
             The corresponding diagrams are depicted in Fig.~\ref{fig2} and worked out in detail in App.~\ref{app:ff-diagrams}.
             The upper left diagram describes the impulse-approximation coupling of the photon to the dressed quark and involves the quark-photon vertex.
             The lower left diagram is the respective coupling to the diquark and depends on the axial-vector diquark-photon vertex.
             The upper right diagram depicts the photon's coupling to the exchanged quark in the quark-diquark kernel, and the lower two diagrams its coupling to the
             diquark amplitudes which involve seagull vertices.

             At the level of the constituents, electromagnetic current conservation $Q^\mu J^{\mu,\rho\sigma}=0$ translates to Ward-Takahashi identities which
             constrain these vertices and relate them to the previously determined quark and diquark propagators and diquark amplitudes.
             Nevertheless, the vertices may involve parts transverse to the photon momentum which are not constrained by current conservation 
	     and
	     yet encode important physics.
             A self-consistent determination of such transverse parts is in principle possible but requires certain numerical effort.
             For instance, the quark-photon vertex can be computed from its rainbow-ladder truncated inhomogeneous Bethe-Salpeter equation
             which unambiguously fixes its transverse contribution~\cite{Maris:1999bh}. As expected from vector-meson dominance models, the latter
             exhibits a $\rho-$meson pole at $Q^2 =-m_\rho^2$.

             In the present calculation we construct the quark-photon vertex from its component fixed by the WTI, i.e. the Ball-Chiu vertex,
             augmented by a transverse $\rho$-meson pole contribution that is modeled after the result in~\cite{Maris:1999bh}.
             An analogous construction is used for the axial-vector seagull vertex. Having fixed those, the axial-vector diquark-photon vertex is completely specified.
             The details of the construction are presented in Apps.~\ref{app:qpv}--\ref{app:dqpv}.

%%%%%%%%%%%%%%%%%%%%%%%%%%%%%%%%%%%%%%%%%%%%%%%%%%%%%%%%%%%%%%%%%%%%%%%%%%%%%%%%%%%%%%%%%%%%%%%%%%%%%%%%%%%%%%%%%%%%%%%%%%%%

         \begin{figure*}[htp]
                    \begin{center}

                    \includegraphics[scale=2.97]{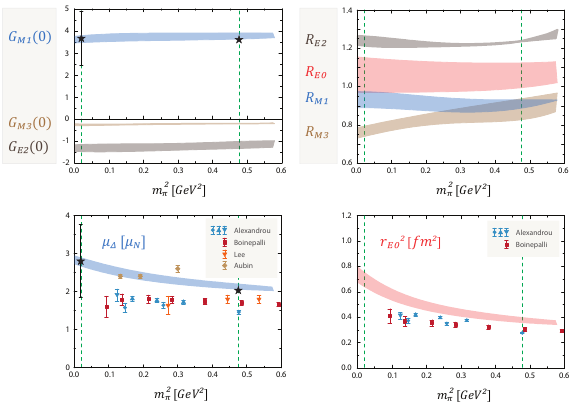}
                    \caption{(Color online) Static electromagnetic properties of the $\Delta$.
                                            \textit{Upper left panel:} dimensionless form factors at $Q^2=0$.
                                            \textit{Upper right panel:} dimensionless squared radii $R_i$, given in (GeV\,fm)$^2$.
                                            \textit{Lower left panel:} $\Delta$ magnetic moment, expressed in static nuclear magnetons
                                                                       and compared to lattice data \cite{Alexandrou:2009hs,Boinepalli:2009sq,Lee:2005ds,Aubin:2008qp}.
                                            \textit{Lower right panel:} squared electric charge radius
                                            compared to lattice results \cite{Alexandrou:2009hs,Boinepalli:2009sq}.
                                            Stars denote experimental values~\cite{Amsler:2008zzb}: for the $\Delta$ we plot $\mu_{\Delta^{++}}/2$; for the $\Omega^-$ we show $|\mu_{\Omega^-}|$.
                                            The dashed vertical lines indicate the positions of the $u/d-$ and strange-quark mass.
                                            Note that there is no $s\bar{s}$ pseudoscalar meson in nature; the value $m_{s\conjg{s}}=0.69$ GeV corresponds to a meson-BSE solution
                                            at a strange-quark mass $m_s = 150$ MeV~\cite{Holl:2004fr}.
                                            }\label{fig:static}

                    \end{center}
        \label{fig.2}
        \end{figure*}

    \renewcommand{\arraystretch}{1.0}

             \begin{table*}[t]

                \begin{center}
                \begin{tabular}{   c @{\;\;} |  @{\;\;}c@{\;\;} || @{\;\;}c@{\;\;} || @{\;\;}c@{\;\;} | @{\;\;}c@{\;\;} | @{\;\;}c@{\;\;} ||  @{\;\;}c@{\;\;} | @{\;\;}c@{\;\;} | @{\;\;}c@{\;\;} | @{\;\;}c@{\;\;} || @{\;\;}c@{\;\;} | @{\;\;}c@{\;\;}  }

                    $m_\pi$        &  $\Lambda$     &     $M_\Delta$  &  $G_{M1}(0)$   &  $G_{E2}(0)$ &  $G_{M3}(0)$ &  $R_{E0}$ &  $R_{M1}$ &  $R_{E2}$ &  $R_{M3}$ &  $\mu_\Delta$ &  $r_{E0}^2$  \\   \hline

                    $0.14$         &  $0.98$        &     $1.73(5)$   &   $3.64(16)$   &  $-1.32(16)$ &  $-0.26(4)$ &  $1.06(9)$ &  $0.93(4)$ &  $1.24(2)$ &  $0.77(3)$ &  $2.78(12)$ &  $0.70(6)$              \\
                    $0.69$         &  $0.76$        &     $1.81(4)$   &   $3.77(13)$   &  $-1.18(16)$ &  $-0.20(2)$ &  $1.06(6)$ &  $0.91(2)$ &  $1.24(1)$ &  $0.88(5)$ &  $2.12(8)$ &  $0.38(2)$

                \end{tabular} \caption{Results for the $\Delta$ mass and static electromagnetic properties at two different pion masses,
                                       where $\Lambda$ is the input scale of Eq.\,\eqref{couplingMT} and the $\eta$ dependence is indicated in parentheses.
                                       $m_\pi$, $\Lambda$ and $M_\Delta$ are given in GeV, $\mu_\Delta$ in nuclear magnetons and $r_{E0}^2$ in fm$^2$.
                                       The $G_i(0)$ are dimensionless and the $R_i$ are given in (GeV\,fm)$^2$.
                                       }\label{tab:results}
                \end{center}

        \end{table*}

\section{Results and discussion} \label{sec:results}

       We have calculated the electromagnetic properties of the $\Delta$
       within the decomposition of Fig.~\ref{fig2} for the electromagnetic current.
       This is achieved by specifying expressions for the quark-photon vertex, the diquark-photon vertices and the seagull terms.
       All necessary ingredients that enter this calculation, such as the quark and
       diquark propagators and the axial-vector diquark and $\Delta$ amplitudes, were explained in Section~\ref{sec:qdq}.
       Through the decomposition of Fig.~\ref{fig2} all resulting $\Delta$ properties are traced back to the effective quark-gluon coupling in Eq.\,\eqref{couplingMT}.
       Specifically, one can investigate the impact of the infrared properties, controlled by the width parameter $\eta$ (cf.~Fig.~\ref{fig:coupling}),
       on resulting observables. This is indicated by the colored bands in Figs.~\ref{fig3} and \ref{fig:static}.

       In Fig.~\ref{fig3} we depict the core contributions to the $\Delta^+$ electromagnetic form factors,
       calculated at the physical point $m_\pi=140$ MeV and compared to lattice data at three different pion masses~\cite{Alexandrou:2009hs}.
       As can be seen from the figure, we find a good overall agreement between our results and those obtained on the lattice.
       Because we assume isospin symmetry, the $\Delta^{++}$, $\Delta^0$ and $\Delta^-$ form factors
       are simply obtained by multiplying those of the $\Delta^{+}$ with the appropriate charges, cf. App.~\ref{app:color-flavor-charge}.
       We note that the $\Delta$ in our core calculation is a stable bound state and not a resonance, i.e. it does not develop a width.
       Non-analyticities associated with the decay channel $\Delta\to N\pi$ which would appear for $M_\Delta-M_N>m_\pi$,
       corresponding to the domain $m_\pi \lesssim 300$ MeV,
       are therefore absent.
       The same is true for the available lattice results which are obtained at pion masses above this threshold.

       Since the $\Delta$ is highly unstable, the available experimental data for its electromagnetic decays are rather poor.
       Only the magnetic moments of $\Delta^{+}$ and $\Delta^{++}$ are experimentally known, albeit with large errors.
       The corresponding values for $G_{M1}(0)$ are $7.3\pm 2.5$ ($\Delta^{++}$) and $3.5^{+7.2}_{-7.6}$ ($\Delta^+$).
       Our result $G_{M1}(0) = 3.64(16)$ compares well with quark-model predictions and chirally extrapolated lattice results
       which typically quote values $G_{M1}(0)\sim3\dots 4$,
       see~\cite{Ramalho:2010xj} and references therein.

       The deformation of the $\Delta$ is encoded in its electric quadrupole and magnetic octupole form factors.
       Non-relativistically, a negative sign for the electric quadrupole moment $G_{E2}(0)$
       indicates an oblate charge distribution for the $\Delta$ in the Breit frame.
       (A different interpretation arises in the infinite-momentum frame, see~\cite{Alexandrou:2009hs}).
       From the measurement of the $N\gamma\Delta$ transition, one can infer
       the value $G_{E2}(0)=-1.87(8)$ in the large-$N_C$ limit~\cite{Buchmann:2002mm,Alexandrou:2009hs};
       comparable values are predicted by a range of constituent-quark models~\cite{Ramalho:2009vc}.
       The impact of pionic corrections upon the low-$Q^2$ behavior of $G_{E2}$ is however unclear.
       While the lattice data of Ref.~\cite{Alexandrou:2009hs} are limited by large statistical errors
       which prevent an accurate extrapolation to $Q^2=0$,
       they indicate a negative value for $G_{E2}(0)$ as well.
       Our result for the electric quadrupole moment, $G_{E2}(0)=-1.32(16)$,
       is negative and compatible with these results.
       We note that $G_{E2}(Q^2)$ develops a zero-crossing at $Q^2/M_\Delta^2\sim 0.6$,
       a feature which is unexpected but not clearly excluded from the available lattice results.

       The lattice signal for the magnetic octupole form factor $G_{M3}(Q^2)$ is weak and
       plagued by large error bars, especially at low $Q^2$.
       At next-to-leading order in a chiral expansion, the magnetic octupole moment vanishes~\cite{Arndt:2003we}.
       Our calculation yields a small and negative form factor;
       the corresponding magnetic octupole moment is $G_{M3}(0)=-0.26(4)$.

        An ambiguity arises when comparing our form factor results, calculated with an implicit 'core' mass $M_\Delta>M_\Delta^\text{exp}$, to experimental and lattice data.
        Form factors are dimensionless, hence they can only depend on dimensionless variables.
        To account for this, the electromagnetic form factors in Fig.\,\ref{fig3} are plotted
        as a function of the dimensionless variable $Q^2/M_\Delta^2$ for our data and $Q^2/(M_\Delta^\text{lat})^2$ for the lattice data.
        From another point of view, $M_\Delta$ defines an effective scale of dynamical chiral symmetry breaking.
        Once such a scale is set (e.g. by having numerically computed $M_\Delta$), all dimensionful quantities
        can be related to this scale. On the lattice, a scale must be defined as well to convert dimensionless lattice results into physical units.
        Hence, an unambiguous comparison between different theoretical approaches and experiment should ideally involve dimensionless quantities.

        In the upper left panel of Fig.\,\ref{fig:static} we show our results for the dimensionless form factors $G_{M1}$, $G_{E2}$ and $G_{E3}$
        at vanishing photon momentum transfer $Q^2=0$.
        For two quark flavors with equal masses the $\Delta^-$ becomes identical to the $\Omega^-$-baryon
        when evaluated at the value for the strange-quark mass, as the $\Omega^-$ is a pure $sss$ state.
        Experimentally, $M_{\Omega}=1.672$ GeV and $\mu_{\Omega^-}=-2.02(5) \mu_N$ which implies
        $|G_{M1}(0)| = M_{\Omega}/M_N\,|\mu_{\Omega^-}|  = 3.59(9)$.
        Our result for $G_{M1}(0)$ is almost independent of the current-quark mass and agrees reasonably well with the experimental value for the $\Omega^-$.
        This is an encouraging result as pionic effects should have diminished in the vicinity of the strange-quark mass above which the baryon is increasingly dominated by its core.
        A similar behavior can also be observed for $G_{E2}(0)$ and $G_{M3}(0)$ which are negative throughout the current-mass range.

        The $\Delta$ magnetic moment is given by
        \begin{equation}\label{magneticmoment}
            \mu_\Delta^\text{dim} = \frac{e}{2M_\Delta}\,G_{M1}(0) = \frac{e}{2M_N^\text{exp}}\left[ G_{M1}(0)\,\frac{M_N^\text{exp}}{M_\Delta}\right],
        \end{equation}
        where $M_\Delta$ is running with the current-quark mass; hence its value in static nuclear magnetons is given by
        the bracket in Eq.\,\eqref{magneticmoment}.
        To compare with experiment or lattice, we must again bear in mind that our calculated $\Delta$ mass is different
        from the one which is measured (or obtained in lattice calculations), and that
        the unambiguous comparison of magnetic moments is that of the dimensionless value $G_{M1}(0)$.
        To account for this, we plot $\mu_\Delta$ in Fig.\,\ref{fig:static} by replacing $M_\Delta$ in Eq.\,\eqref{magneticmoment} by
        the following reference mass:
        \begin{equation}\label{referencemass}
           M_\Delta^\text{Ref}(m_\pi^2)^2 = M_0^2 + \left(\frac{3 m_\pi}{2}\right)^2 \left(1 + f(m_\pi^2)\right) ,
        \end{equation}
        with $f(m_\pi^2) = 0.77/(1+(m_\pi/0.59\,\text{GeV})^4)$ and $M_0=1.2$ GeV.
        Eq.\,\eqref{referencemass} reproduces the experimental $\Delta$ and $\Omega^-$ masses at $m_\pi=0.14$ GeV
        and $m_\pi=0.69$ GeV, respectively, and provides a reasonable representation
        of the dynamical lattice results of Ref.\,\cite{Alexandrou:2009hs} for $M_\Delta$.
        At the $u/d-$quark mass, our calculated value for the magnetic moment is thus $\mu_\Delta = 2.78(12) \mu_N$.
        Because of the constancy of $G_{M1}(0)$ it decreases with an inverse power of the mass $M_\Delta^\text{Ref}(m_\pi^2)$.

        The charge radius corresponding to a form factor $G_i(Q^2)$ is defined as
        \begin{equation}
           r_i^2 = -\frac{6}{G_i(0)} \left.\frac{dG_i}{dQ^2}\right|_{Q^2=0}\,(\hbar c)^2\,,
        \end{equation}
        where $\hbar c =0.197$ GeV fm.
        We display the dimensionless squared charge radii $R_i := r_i^2 \, M_\Delta^2 $ in the upper right panel
        of Fig.\,\ref{fig:static}. Again their current-mass dependence is found to be weak.
        The dependence on the infrared width $\eta$ arises from $M_\Delta$ which has a sizeable dependence on $\eta$, cf. Table~\ref{tab:results}.
        In the lower right panel of Fig.~\ref{fig:static} we display the rescaled squared charge radius
        \begin{equation}\label{radii_rescaled}
            r_i^2\,\frac{M_\Delta^2}{\left(M_\Delta^\text{Ref}\right)^2} = \frac{R_{E0}}{\left(M_\Delta^\text{Ref}\right)^2}\,.
        \end{equation}
        It shows a satisfactory agreement with the lattice data at larger pion masses, where the $\Delta$ is mainly described by its core.
        At the $u/d$ mass, its value is $0.70(6)$ fm$^2$.

        It is noteworthy that an investigation of dimensionless properties
        effectively removes the model dependence between the 'core' version and the fixed-scale version of the coupling in Eq.\,\eqref{couplingMT}
        which was discussed in Section~\ref{sec:qprop}.
        In the chiral limit, in the absence of a current-quark or pion mass, the input scale $\Lambda$ of Eq.\,\eqref{couplingMT}
        is the only relevant scale in all our calculations. For instance, hadron masses scale with $\Lambda$.
        Since the properties of a quark core were modeled solely by increasing $\Lambda$,
        the discrepancy between the two models disappears when comparing dimensionless values.
        Nevertheless, since both models describe a quark core (pionic effects are absent in either of them),
        characteristic chiral features induced by pionic corrections, such as a logarithmically divergent $\Delta$ charge radius, can naturally not be reproduced.

        On a related note, the estimation of chiral cloud effects from a pure quark-core calculation is not unambiguous.
        For instance, adding pionic effects in the quark-diquark kernel would affect the Bethe-Salpeter normalization of the $\Delta$ amplitude,
        a condition which is equivalent to charge conservation: $G_{E0}(0)=1$. Charge conservation must always be satisfied, irrespective
        of whether pionic effects are included or not.
        Hence the effective core contribution to the form factors of a $\Delta$-baryon that is dressed by pionic degrees of freedom would be smaller.
        The resulting 'renormalization' of all form factors would be independent of $Q^2$ but would depend on the current-quark mass: at large masses,
        the difference should diminish as pionic effects become small.
        It is reassuring that the $\Delta$ magnetic moment and electric charge radius displayed in Fig.~\ref{fig:static}
        agree quite well with experimental and lattice results in this domain.

        We have established that the static limits of the four form factors depend only weakly on the current-quark mass.
        It is remarkable that the same can also be said for their $Q^2-$dependence:
        when plotted over $Q^2/M_\Delta^2$, the overall shape of Fig.\,\ref{fig3} persists throughout the current-quark mass range.
        For instance, the position of the zero crossing in $G_{E2}(Q^2)$, expressed through $Q^2/M_\Delta^2$, is almost independent of the value of $m_\pi$.
        Thus, our resulting form factors for the $\Omega^-$ are not materially different from those of the $\Delta$ once their inherent mass dependence is scaled out.
        The same pattern appears to be valid for the lattice results as well: when plotted over $Q^2/M_\Delta^2$,
        the results at different pion masses follow a similar $Q^2-$evolution.

       We finally note that the current setup prohibits an investigation of form factors beyond $Q^2/M_\Delta^2 \gtrsim 2$
       because of timelike and/or complex singularities that appear in the quark propagator and quark-quark scattering matrix.
       Irrespective of its details which are truncation-dependent, such a singularity structure is an inevitable feature of these Green functions
       and must be accounted for in order to investigate the large-$Q^2$ domain.
       As an intermediate step it might be worthwhile to study this regime with the help of
       pole-free model propagators,
       a procedure which has been adopted in Ref.~\cite{Cloet:2008re} in the context of nucleon electromagnetic form factors.

%%%%%%%%%%%%%%%%%%%%%%%%%%%%%%%%%%%%%%%%%%%%%%%%%%%%%%%%%%%%%%%%%%%%%%%%%%%%%%%%%%%%%%%%%%%%%%%%%%%%%%%%%%%%%%%%%%%%%%%%%%%%
\section{Conclusions}

      We provided a calculation of $\Delta$ 
      and the $\Omega$ 
      electromagnetic form factors
      in a Poincar´e-covariant quark-diquark approach.
      All quark and diquark ingredients were determined self-consistently from
      Dyson-Schwinger and Bethe-Salpeter equations and thereby related to the
      fundamental quantities in QCD.
      We employed a rainbow-ladder truncation which corresponds to a dressed gluon exchange between
      the quarks inside the diquark.
      Since pion-cloud effects are not implemented, our results describe the $\Delta$ quark core.

      The $Q^2-$evolution of the electromagnetic form factors agrees well with results from lattice QCD.
      In particular, the electric quadrupole and magnetic octupole form factors are negative throughout the
      current-quark mass range
      which, in the traditional interpretation, indicates an oblate deformation of the $\Delta$'s charge and magnetization distributions.
      The dimensionless form factors at vanishing photon-momentum transfer are almost independent of the current-quark mass.
      At larger quark masses, where pionic effects do not contribute anymore,
      our results for the $\Delta$ magnetic moment and electric charge radius are close to those obtained in lattice QCD.
      Moreover, the magnetic moment evaluated at the strange-quark mass agrees well with the experimental value for the $\Omega^-$.

      In summary, the results collected herein show good agreement with lattice observations,
      quark model analyses and the available experimental data.
      Near-future measurements at MAMI and JLab facilities remain to validate these predictions for the $\Delta$'s electromagnetic properties.
      From this perspective, while the $N\to\Delta\gamma$ transition is accurately measured,
      a forthcoming extension of our approach to the investigation of spin-$\nicefrac{3}{2} \rightarrow$ spin-$\nicefrac{1}{2}$
      electromagnetic transitions will constitute a nontrivial test for the Poincar\'{e}-covariant framework presented herein.
      Moreover, a study of the evolution of the $\Delta$ electromagnetic form factors at large values of the photon momentum transfer is desirable.
      Naturally this constitutes a very challenging task, in particular since data collection for electromagnetic properties of baryon resonances at large $Q^2$ is due to begin.

      Finally, in view of a better understanding of the structure of baryons,
      our approach can be methodically improved by eliminating the diquark ansatz in support of a fully Poincar\'{e}-covariant solution
      of the three-quark Faddeev equation. We have recently reported on numerical results for the nucleon mass in such a framework 
      \cite{Eichmann:2009qa}.
      In parallel we intend to augment our analysis by developing compatible tools to incorporate missing chiral cloud effects.

%%%%%%%%%%%%%%%%%%%%%%%%%%%%%%%%%%%%%%%%%%%%%%%%%%%%%%%%%%%%%%%%%%%%%%%%%%%%%%%%%%%%%%%%%%%%%%%%%%%%%%%%%%%%%%%%%%%%%%%%%%%%

     \section{Acknowledgements}
     D. Nicmorus thanks D. Rischke for his support. We would also like to thank
     I. C. Clo\"et, C. S. Fischer, A. Krassnigg, G. Ramalho, and R. Williams
     for fruitful discussions. This work was supported by the Austrian Science Fund FWF under
     Project No.~P20592-N16 and Erwin-Schr\"odinger-Stipendium No.~J3039,
     by the Helmholtz Young Investigator Grant VH-NG-332, and
     by the Helmholtz  International Center for FAIR within the framework of the LOEWE program launched by the State of Hesse,
     GSI, BMBF and DESY.

%%%%%%%%%%%%%%%%%%%%%%%%%%%%%%%%%%%%%%%%%%%%%%%%%%%%%%%%%%%%%%%%%%%%%%%%%%%%%%%%%%%%%%%%%%%%%%%%%%%%%%%%%%%%%%%%%%%%%%%%%%%%%%%%%%%%%%%%%%%%%%%%%%%%%%%%%%%%%%%%%%%%%%%%%%%%%
%%%%%%%%%%%%%%%%%%%%%%%%%%%%%%%%%%%%%%%%%%%%%%%%%%%%%%%%%%%%%%%%%%%%%%%%%%%%%%%%%%%%%%%%%%%%%%%%%%%%%%%%%%%%%%%%%%%%%%%%%%%%%%%%%%%%%%%%%%%%%%%%%%%%%%%%%%%%%%%%%%%%%%%%%%%%%
%%%%%%%%%%%%%%%%%%%%%%%%%%%%%%%%%%%%%%%%%%%%%%%%%%%%%%%%%%%%%%%%%%%%%%%%%%%%%%%%%%%%%%%%%%%%%%%%%%%%%%%%%%%%%%%%%%%%%%%%%%%%%%%%%%%%%%%%%%%%%%%%%%%%%%%%%%%%%%%%%%%%%%%%%%%%%

\begin{appendix}

  \section{Euclidean conventions} \label{app:euclidean}

            We work in Euclidean momentum space with the following conventions:
            \begin{equation}
                p\cdot q = \sum_{k=1}^4 p_k \, q_k,\quad
                p^2 = p\cdot p,\quad
                \Slash{p} = p\cdot\gamma\,.
            \end{equation}
            A vector $p$ is spacelike if $p^2 > 0$ and timelike if $p^2<0$.
            The hermitian $\gamma-$matrices $\gamma^\mu = (\gamma^\mu)^\dag$ satisfy the anticommutation relations
            $\left\{ \gamma^\mu, \gamma^\nu \right\} = 2\,\delta^{\,\mu\nu}$, and we define
            \begin{equation}
                \sigma^{\mu\nu} = -\frac{i}{2} \left[ \gamma^\mu, \gamma^\nu \right]\,, \quad
                \gamma^5 = -\gamma^1 \gamma^2 \gamma^3 \gamma^4\,.
            \end{equation}
            In the standard representation one has:
            \begin{equation*}
                \gamma^k  =  \left( \begin{array}{cc} 0 & -i \sigma^k \\ i \sigma^k & 0 \end{array} \right), \;
                \gamma^4  =  \left( \begin{array}{c@{\quad}c} \mathds{1} & 0 \\ 0 & \!\!-\mathds{1} \end{array} \right), \;
                \gamma^5  =  \left( \begin{array}{c@{\quad}c} 0 & \mathds{1} \\ \mathds{1} & 0 \end{array} \right),
            \end{equation*}
            where $\sigma^k$ are the three Pauli matrices.
            The charge conjugation matrix is given by
            \begin{equation}
                C = \gamma^4 \gamma^2, \quad C^T = C^\dag = C^{-1} = -C\,,
            \end{equation}
            and the charge conjugates for (pseudo-)\,scalar, \mbox{(axial-)} vector and tensor amplitudes are defined as
            \begin{equation}
            \begin{split}
                \conjg{\Gamma}(p,P) &:= C\,\Gamma(-p,-P)^T\,C^T \,,   \\
                \conjg{\Gamma}^\alpha(p,P) &:= -C\,{\Gamma^\alpha}(-p,-P)^T\,C^T \,,   \\
                \conjg{\Gamma}^{\beta\alpha}(p,P) &:= C\,{\Gamma^{\alpha\beta}}(-p,-P)^T\,C^T\,,
            \end{split}
            \end{equation}
            where $T$ denotes a Dirac transpose.
            Four-momenta are conveniently expressed through hyperspherical coordinates:
            \begin{equation}\label{APP:momentum-coordinates}
                p^\mu = \sqrt{p^2} \left( \begin{array}{l} \sqrt{1-z^2}\,\sqrt{1-y^2}\,\sin{\phi} \\
                                                           \sqrt{1-z^2}\,\sqrt{1-y^2}\,\cos{\phi} \\
                                                           \sqrt{1-z^2}\;\;y \\
                                                           \;\; z
                                         \end{array}\right),
            \end{equation}
            and a four-momentum integration reads:
            \begin{equation*} \label{hypersphericalintegral}
                 \int \!\!\frac{d^4 p}{(2\pi)^4} = \frac{1}{(2\pi)^4}\,\frac{1}{2}\int\limits_0^{\infty} dp^2 \,p^2 \int\limits_{-1}^1 dz\,\sqrt{1-z^2}  \int\limits_{-1}^1 dy \int\limits_0^{2\pi} d\phi \,.
            \end{equation*}

  \section{$\Delta$ Electromagnetic current} \label{sec:current}

        \subsection{General properties}

             The matrix-valued electromagnetic current of the $\Delta$ can be written in the most general form as
             \begin{equation}\label{eq:current-general1}
                 J^{\mu,\rho\sigma}(P,Q) = \mathds{P}^{\rho\alpha}(P_f)\,\Gamma^{\mu,\alpha\beta}(P,Q)\,\mathds{P}^{\beta\sigma}(P_i)\,.
             \end{equation}
             The vertex involves two momenta, expressed through the ingoing and outgoing momenta $P_i$, $P_f$ or by the
             average momentum $P=(P_i+P_f)/2$ and photon momentum $Q=P_f-P_i$. Since the particle is on-shell, $P_i^2=P_f^2=-M_\Delta^2$, one has
             \begin{equation}
                 P^2 = -M_\Delta^2 (1+\tau), \qquad P\cdot Q=0\,,
             \end{equation}
             where $\tau := Q^2/(4M_\Delta^2)$; hence the Lorentz-invariant form factors which constitute the vertex
             can only depend on the photon momentum-transfer $Q^2$.
             The Rarita-Schwinger projector $\mathds{P}^{\alpha\beta}(K)$ for a general momentum $K$ is defined via
             \begin{equation}\label{eq:RSprojector1}
                 \mathds{P}^{\alpha\beta}(K) := \Lambda_+(K) \left( T^{\alpha\beta}_K - \frac{1}{3}\,\gamma_T^\alpha\,\gamma_T^\beta\right),
             \end{equation}
             with the transverse projector $T^{\alpha\beta}_K = \delta^{\alpha\beta} - \hat{K}^\alpha \hat{K}^\beta$,
             the transverse gamma matrices $\gamma_T^\alpha = T^{\alpha\beta}_K \gamma^\beta$, and
             the positive-energy projector $\Lambda_+(K) =(\mathds{1} + \hat{\Slash{K}})/2$. $\hat{K}$ denotes a normalized 4-vector.
             The projectors $\Lambda_+(K)$ and $\mathds{P}^{\alpha\beta}(K)$ satisfy the relations
             \begin{equation}\label{eq:RSrelations}
             \begin{split}
                 &\Lambda_+(K)\,\Slash{\hat{K}} = \Slash{\hat{K}}\,\Lambda_+(K) = \Lambda_+(K)\,, \\
                 &\mathds{P}^{\alpha\beta}(K)\,K^\beta = K^\alpha \mathds{P}^{\alpha\beta}(K) = 0\,, \\
                 &\mathds{P}^{\alpha\beta}(K)\,\gamma^\beta = \gamma^\alpha \mathds{P}^{\alpha\beta}(K) = 0\,.
             \end{split}
             \end{equation}

             By contracting Eq.\,\eqref{eq:current-general1} with Rarita-Schwinger spinors $\conjg{u}^\rho(P_f,s_f)$, $u^\sigma(P_i,s_i)$
             which are eigenstates of $\mathds{P}^{\alpha\beta}$:
             \begin{equation}
                 \mathds{P}^{\alpha\beta}(K) u^\beta(K,s) = u^\alpha(K,s)\,, \quad
                 s \in\{ \pm\nicefrac{3}{2},\pm\nicefrac{1}{2}\}\,,
             \end{equation}
            we obtain the current-matrix element
             $\langle P_f, s_f \,|\, J^\mu\,| \,P_i, s_i \rangle$.

             The positive-energy projector is invariant under charge conjugation:
             \begin{equation}
                 \conjg{\Lambda}_+(K) := C\,\Lambda_+(-K)^T\,C^T = \Lambda_+(K)\,,
             \end{equation}
             since $C\,C^T = 1$ and $C\,{\gamma^\mu}^T = -\gamma^\mu\,C$. The same is true for the
             charge-conjugated Rarita-Schwinger projector:
             \begin{eqnarray}
                 \conjg{\mathds{P}}^{\beta\alpha}(K) :&=& C\,\mathds{P}^{\alpha\beta}(-K)^T\,C^T = \nonumber \\
                                                   &=& \left( T^{\beta\alpha}_K - \frac{1}{3}\,\gamma_T^\beta\,\gamma_T^\alpha\right)\Lambda_+(K) = \\
                                                   &=& \Lambda_+(K) \left( T^{\beta\alpha}_K - \frac{1}{3}\,\gamma_T^\beta\,\gamma_T^\alpha\right) = \mathds{P}^{\beta\alpha}(K)\,.  \nonumber
             \end{eqnarray}
             $\Lambda_+(K)$ commutes with the bracket since $\gamma^\alpha_T\,\Lambda_+(K) = \Lambda_-(K)\,\gamma^\alpha_T$
             and hence $\gamma_T^\beta \,\gamma_T^\alpha\,\Lambda_+(K) = \Lambda_+(K) \,\gamma_T^\beta \,\gamma_T^\alpha$,
             where $\Lambda_-(K) =(\mathds{1} - \hat{\Slash{K}})/2$ is the negative-energy projector.
             These relations will be useful in the following subsection.

        \subsection{Derivation of the electromagnetic current}\label{app:currentderivation}

             The current $J^{\mu,\rho\sigma}(P,Q)$ in Eq.\,\eqref{eq:current-general1} consists of four form factors.
             $\Gamma^{\mu,\alpha\beta}(P,Q)$ is a five-point function: it has three vector legs and two spinor legs, and it depends on two momenta.
             Such a vertex has 144 tensor structures of definite parity which are given by all combinations of the 36 elements
             \begin{equation} \label{gammastructures}
                 \begin{array}{l}  \gamma^\mu \gamma^\alpha \gamma^\beta \\[0.4cm]
                                   \delta^{\mu\alpha} \gamma^\beta \\ \delta^{\mu\beta} \gamma^\alpha \\ \delta^{\alpha\beta} \gamma^\mu  \end{array}\quad
                 \begin{array}{l}  \gamma^\mu \gamma^\alpha P^\beta \\ \gamma^\mu \gamma^\beta P^\alpha \\ \gamma^\alpha \gamma^\beta P^\mu \\[0.2cm]
                                   \gamma^\mu \gamma^\alpha Q^\beta \\ \gamma^\mu \gamma^\beta Q^\alpha \\ \gamma^\alpha \gamma^\beta Q^\mu  \end{array}\quad
                 \begin{array}{l}  \gamma^\mu P^\alpha P^\beta \\ \gamma^\alpha P^\mu P^\beta  \\ \gamma^\beta P^\alpha P^\mu \\[0.2cm]
                                   \gamma^\mu Q^\alpha Q^\beta \\ \gamma^\alpha Q^\mu Q^\beta  \\ \gamma^\beta Q^\alpha Q^\mu \end{array}\quad
                 \begin{array}{l}  \gamma^\mu P^\alpha Q^\beta \\ \gamma^\mu P^\beta Q^\alpha \\ \gamma^\alpha P^\mu Q^\beta  \\[0.2cm]
                                   \gamma^\alpha P^\beta Q^\mu  \\ \gamma^\beta P^\mu Q^\alpha \\ \gamma^\beta P^\alpha Q^\mu \end{array}
             \end{equation}
             and
             \begin{equation}
                 \begin{array}{l}  Q^\mu Q^\alpha Q^\beta \\ P^\mu P^\alpha P^\beta \end{array}\quad
                 \begin{array}{l}  Q^\mu P^\alpha Q^\beta \\ Q^\mu Q^\alpha P^\beta \\ P^\mu Q^\alpha Q^\beta \\[0.2cm]
                                   Q^\mu P^\alpha P^\beta \\ P^\mu Q^\alpha P^\beta \\ P^\mu P^\alpha Q^\beta \end{array}\quad
                 \begin{array}{l}  \delta^{\mu\alpha} Q^\beta \\ \delta^{\mu\beta} Q^\alpha \\ \delta^{\alpha\beta} Q^\mu \\[0.2cm]
                                   \delta^{\mu\alpha} P^\beta \\ \delta^{\mu\beta} P^\alpha \\ \delta^{\alpha\beta} P^\mu  \end{array}
             \end{equation}
             with the four basic spinor structures
             \begin{equation} \label{spinorstructures}
                 \left\{ \, \mathds{1},\, \Slash{P},\, \Slash{Q},\, \left[\Slash{P}, \Slash{Q}\right] \, \right\}.
             \end{equation}

             Applying the Rarita-Schwinger projectors reduces this set to 8 basis elements.
             First, only the unit matrix $\mathds{1}$ survives upon sandwiching Eq.\,\eqref{spinorstructures} between the positive-energy projectors, since
             for instance
             \begin{equation*}
             \begin{split}
                 \Lambda_+^f \,\Slash{P} \,\Lambda_+^i &= \Lambda_+^f \,\frac{\Slash{P_f}+\Slash{P_i}}{2} \,\Lambda_+^i =  iM \Lambda_+^f \,\Lambda_+^i\,, \\[0.1cm]
                 \Lambda_+^f \,\Slash{Q} \,\Lambda_+^i &= \Lambda_+^f \left(\Slash{P_f}-\Slash{P_i}\right)\Lambda_+^i = 0\,,
             \end{split}
             \end{equation*}
             where we have abbreviated $\Lambda_+^{i,f}=\Lambda_+(P_{i,f})$.
             Moreover, a Rarita-Schwinger contraction of those basis elements in Eq.\,\eqref{gammastructures} which involve instances of $\gamma^\alpha$ or $\gamma^\beta$
             yields zero due to Eq.\,\eqref{eq:RSrelations}, while
             a contraction with $Q^\alpha$ or $Q^\beta$ can be reduced to a contraction with $P^\alpha$ or $P^\beta$ since
             \begin{equation*}
             \begin{split}
                 P^\alpha  \mathds{P}^{\alpha\beta}(P_i) &= \frac{(P_i+P_f)^\alpha}{2}\, \mathds{P}^{\alpha\beta}(P_i)  \stackrel{\eqref{eq:RSrelations}}{=}
                                                            \frac{P_f^\alpha}{2}\,\mathds{P}^{\alpha\beta}(P_i) = \\
                                                      &=       \left(\frac{P^\alpha}{2}+\frac{Q^\alpha}{4}\right) \mathds{P}^{\alpha\beta}(P_i) \\
                 & \Rightarrow \quad  P^\alpha  \mathds{P}^{\alpha\beta}(P_i) = \frac{Q^\alpha}{2}\,\mathds{P}^{\alpha\beta}(P_i)\,.
             \end{split}
             \end{equation*}
             Discarding all basis elements involving $\gamma^\alpha$, $\gamma^\beta$, $Q^\alpha$ or $Q^\beta$ one then
             arrives at the following 8 tensor structures:
             \begin{equation}
                 \begin{array}{l}  \gamma^\mu \delta^{\alpha\beta} \\ P^\mu \delta^{\alpha\beta} \\ Q^\mu \delta^{\alpha\beta} \end{array}\quad
                 \begin{array}{l}  \gamma^\mu Q^\alpha Q^\beta \\ P^\mu Q^\alpha Q^\beta  \\ Q^\mu Q^\alpha Q^\beta \end{array}\quad
                 \begin{array}{l}  \delta^{\mu\alpha} Q^\beta \\ \delta^{\mu\beta} Q^\alpha\,. \end{array}
             \end{equation}

             A further reduction can be achieved by imposing charge-conjugation invariance of the current $J^{\mu,\rho\sigma}(P,Q)$
             which, because the Rarita-Schwinger projectors are charge-conjugation invariant, translates into charge-conjugation invariance of the vertex $\Gamma^{\mu,\alpha\beta}$:
             \begin{equation*}
                 \conjg{\Gamma}^{\mu,\beta\alpha}(P,Q) := -C\,\Gamma^{\mu,\alpha\beta}(-P,-Q)^T\,C^T \stackrel{!}{=} \Gamma^{\mu,\beta\alpha}(P,-Q)\,.
             \end{equation*}
             It reduces the 8 components to 5:
             \begin{equation}
                 \begin{array}{l}  \gamma^\mu \,\delta^{\alpha\beta} \\ P^\mu \,\delta^{\alpha\beta}   \end{array}\quad
                 \begin{array}{l}  \gamma^\mu \,Q^\alpha Q^\beta \\ P^\mu \,Q^\alpha  Q^\beta    \end{array}\quad
                  \delta^{\mu\alpha}  Q^\beta - \delta^{\mu\beta}  Q^\alpha\,.
             \end{equation}
             The resulting current is now automatically conserved, i.e. $Q^\mu J^{\mu,\rho\sigma} = 0$.
             The fifth tensor structure can be related to the first three by the following identity~\cite{Nozawa:1990gt}:
             \begin{equation}
             \begin{split}
                 & \quad \frac{i}{2M} \,\Lambda_+^f  \left[ \delta^{\mu\alpha}  Q^\beta - \delta^{\mu\beta}  Q^\alpha \right]\Lambda_+^i = \\
                 &= \Lambda_+^f \left[ \left( (1+\tau)\,\gamma^\mu + \frac{i P^\mu}{M}\right) \delta^{\alpha\beta} -  \gamma^\mu\,\frac{Q^\alpha Q^\beta}{2M^2}\right] \Lambda_+^i.
             \end{split}
             \end{equation}

             As a consequence, the general electromagnetic current of the $\Delta$ depends on four form factors $F_i^\star(Q^2)$ and can be written as:
             \begin{equation}
                 J^{\mu,\rho\sigma}(P,Q) = \mathds{P}^{\rho\alpha}(P_f)\,\Gamma^{\mu,\alpha\beta}(P,Q)\,\mathds{P}^{\beta\sigma}(P_i)\,,
             \end{equation}
             with
             \begin{equation}\label{eq:current1}
             \begin{split}
                  \Gamma^{\mu,\alpha\beta}(P,Q) &= \left( (F_1^\star + F_2^\star)\,i\gamma^\mu - F_2^\star\,\frac{P^\mu}{M}\right)\delta^{\alpha\beta}    \\
                          & - \left( (F_3^\star+F_4^\star)\,i\gamma^\mu - F_4^\star\,\frac{P^\mu}{M}\right) \frac{Q^\alpha Q^\beta}{4M^2}\,.
             \end{split}
             \end{equation}
             Using the Gordon identity
             \begin{equation}\label{eq:gordonid}
                 \Lambda_+^f \left[ \gamma^\mu + \frac{iP^\mu}{M} + \frac{\sigma^{\mu\nu}Q^\nu}{2M}\right]\Lambda_+^i = 0
             \end{equation}
             finally leads to the expression given in Eq.\,\eqref{eq:current2}.

    \renewcommand{\arraystretch}{1.2}

        \subsection{Extraction of the form factors}

              The electromagnetic form factors are extracted from Lorentz-invariant traces of the current $J^{\mu,\alpha\beta}$.
              Each index $\mu,\alpha,\beta$ can be contracted with the momenta $P$, $Q$ or a gamma matrix.
              For the Rarita-Schwinger indices $\alpha$, $\beta$
              it is sufficient to consider the contraction with the momentum $P$ due to the relations
              \begin{equation}
                  P^\alpha  \mathds{P}^{\alpha\beta}(P_i) = \frac{Q^\alpha}{2}\,\mathds{P}^{\alpha\beta}(P_i)\,, \quad
                  \gamma^\alpha\,\mathds{P}^{\alpha\beta}(k) = 0\,.
              \end{equation}
              The vector index $\mu$ can be contracted with either $P^\mu$ and $\gamma^\mu$; a contraction with $Q^\mu$ yields zero
              because of current conservation. In addition, all indices can be contracted among themselves.
              Upon performing the Dirac traces one arrives at the following Lorentz-invariant scalars:
              \begin{equation} \label{sses}
              \begin{array}{l} s_1 := \text{Tr}\left\{ J^{\mu,\alpha\beta} \right\} \hat{P}^\mu  \hat{P}^\alpha  \hat{P}^\beta  \\
                               s_2 := \text{Tr}\left\{ J^{\mu,\alpha\alpha} \right\} \hat{P}^\mu \\[0.2cm]
                               s_3 := \text{Tr}\left\{ J^{\mu,\alpha\beta} \,\gamma^\mu_T  \right\} \hat{P}^\alpha  \hat{P}^\beta \\
                               s_4 := \text{Tr}\left\{ J^{\mu,\alpha\alpha} \,\gamma^\mu_T \right\} \,.
              \end{array}
              \end{equation}
              Further possibilities such as $\text{Tr} \{ J^{\mu,\alpha\beta} \,\Slash{\hat{P}}  \}\hat{P}^\mu  \hat{P}^\alpha  \hat{P}^\beta $
              or $\text{Tr}\left\{ J^{\mu,\mu\alpha} \right\}\hat{P}^\alpha=\text{Tr}\left\{ J^{\mu,\alpha\mu} \right\}\hat{P}^\alpha$
              linearly depend on \eqref{sses}. We used $\hat{P}=P/(iM)$ to arrive at dimensionless quantities and $\gamma^\mu_T := T^{\mu\nu}_P \gamma^\nu$
              to project out the component in $\hat{P}-$direction. A computation in the Breit frame, where
              \begin{equation}\label{breitframe}
                  \begin{array}{l} P = iM\sqrt{1+\tau} \,\vect{e}_4\,, \\ Q = 2M\sqrt{\tau}\,\vect{e}_3  \end{array} \quad \Rightarrow \quad
                  \begin{array}{l} \hat{P} = \vect{e}_4\,, \\ \hat{Q} = \vect{e}_3\,, \end{array}
              \end{equation}
              yields the result:
              \begin{align*}
                  s_1 &= \frac{4i\,\tau}{9} \,\sqrt{1+\tau} \left[ \,3\, G_{E_0} + 2\,\tau\,G_{E_2} \,\right],  \\
                  s_2 &= 4i \,\sqrt{1+\tau} \left[ \left(1+\frac{2\,\tau}{3}\right) G_{E_0} + \frac{4\tau^2}{9}\,G_{E_2} \,\right],  \\
                  s_3 &= -\frac{16 i\,\tau^2}{9}  \left[ G_{M_1} + \frac{6\,\tau}{5}\,G_{M_3} \,\right],  \\
                  s_4 &= -\frac{8 i\,\tau}{9}   \left[ (4\,\tau+5)\, G_{M_1} + \frac{24\,\tau^2}{5}\,G_{M_3} \,\right].
              \end{align*}
              Since the $s_i$ are Lorentz-invariant, this result is frame-independent.
              Its inversion yields for the electric form factors
              \begin{equation}
              \begin{split}
                  G_{E_0} &= \frac{s_2-2s_1}{4i\sqrt{1+\tau}}\,, \\
                  G_{E_2} &= \frac{3}{8i\,\tau^2\sqrt{1+\tau}} \left[ 2s_1 \left(\tau+\frac{3}{2}\right) - \tau s_2 \right],
              \end{split}
              \end{equation}
              and for the magnetic form factors:
              \begin{equation}
              \begin{split}
                  G_{M_1} &= \frac{9i}{40\,\tau}\left(s_4-2s_3\right)\,, \\
                  G_{M_3} &= \frac{3i}{16\,\tau^3} \left[ 2s_3 \left(\tau+\frac{5}{4}\right) - \tau s_4 \right].
              \end{split}
              \end{equation}

  \section{Diagrams in the quark-diquark model} \label{app:ff-diagrams}

    \subsection{General decomposition} \label{app:currentdiagrams}

            In this appendix we collect the ingredients of the $\Delta$ electromagnetic current operator
            in the quark-diquark model which are depicted in Fig.~\ref{fig2}.
            The explicit form of the current is given by a sum of impulse-approximation diagrams (left panel in Fig.~\ref{fig2})
            and two-loop contributions which represent the photon's coupling to the quark-diquark kernel (right panel of Fig.~\ref{fig2}):
            \begin{align}\label{ff:current2}
                J^{\mu,\rho\sigma}  &= \int    \conjg{\Phi}^{\rho\alpha}(p_f,P_f)\, (X_\text{q}+X_\text{dq})^{\mu,\alpha\beta}\, \Phi^{\beta\sigma}(p_i,P_i) \,+ \nonumber \\
                                    & + \int\!\!\!\int  \conjg{\Phi}^{\rho\alpha}(p_f,P_f)\, X_\text{K}^{\mu,\alpha\beta}\, \Phi^{\beta\sigma}(p_i,P_i) \,.
            \end{align}
            Here, $P_i$ and $P_f=P_i+Q$ are incoming and outgoing on-shell $\Delta$ momenta.
            The relative momenta $p_i$ and $p_f$ are independent loop momenta in the two-loop diagrams;
            in the one-loop diagrams they are related to each other:
            $p_f-p_i=(1-\xi)\,Q$ for the quark diagram and $p_f-p_i=-\xi\,Q$ for the diquark diagram,
            where $\xi \in [0,1]$ is an arbitrary momentum-partitioning parameter which must be specified prior to solving the BSE.
            $\alpha,\beta=1\dots 4$ are the diquark's Lorentz indices. The quark-diquark amplitudes $\Phi^{\rho\alpha}$ are the solutions
            of the quark-diquark Bethe-Salpeter equation \eqref{deltabse}.
            The ingredients of Eq.\,\eqref{ff:current2} are given by
            \begin{align}
                X_\text{q}^{\mu,\alpha\beta}  &=  S(p_+)\,\Gamma^\mu_\text{q}(p_+,p_-)\, S(p_-) \, D^{\alpha\beta}(k_-) \,,  \\
                X_\text{dq}^{\mu,\alpha\beta} &=  S(p_-) \, D^{\alpha\alpha'}(k_+) \, \Gamma^{\mu,\alpha'\beta'}_\text{dq}(k_+,k_-) \, D^{\beta'\beta}(k_-) \,, \nonumber \\
                X_\text{K}^{\mu,\alpha\beta}  &= D^{\alpha\alpha'}(k_+)\, S(p_+)\,K^{\mu,\alpha'\beta'} S(p_-) \,D^{\beta'\beta}(k_-)\,.  \nonumber
            \end{align}
            and depend on the quark-photon vertex $\Gamma^\mu_\text{q}$ and the axial-vector diquark photon vertex $\Gamma^{\mu,\alpha\beta}_\text{dq}$.
            The quark and diquark momenta are:
            \begin{align*}
                p_- &= p_i+\xi\,P_i\,,    &   k_- &= -p_i + (1-\xi)\,P_i\;,  \\
                p_+ &= p_f+\xi\,P_f\,,    &   k_+ &= -p_f + (1-\xi)\,P_f\;.
            \end{align*}
            The gauged kernel $K^{\mu,\alpha\beta}$ contains the exchange-quark diagram and the seagull vertex $M^{\mu,\alpha}$:
            \begin{equation}
                K^{\mu,\alpha\beta} = \left( K_\text{EX} + K_\text{SG} + K_{\overline{\text{SG}}} \right)^{\mu,\alpha\beta}\;,
            \end{equation}
            with
            \begin{align}
                K_\text{EX}^{\mu,\alpha\beta}              \!&= \Gamma^\beta(r_+,k_-)\Big[ S(q_+) \,\Gamma^\mu_\text{q}(q_+,q_-)\, S(q_-) \Big]^T \conjg{\Gamma}^\alpha(r_-,k_+)   \nonumber \\
                K_\text{SG}^{\mu,\alpha\beta}              \!&= M^{\mu,\beta}(r'_+,k_-,Q)\, S(q_+)^T\,\conjg{\Gamma}^\alpha(r_-,k_+)\,, \nonumber\\
                K_{\overline{\text{SG}}}^{\mu,\alpha\beta} \!&= \Gamma^\beta(r_+,k_-)\,S(q_-)^T\,\conjg{M}^{\mu,\alpha}(r'_-,k_+,Q)\,,
            \end{align}
            and momenta:
            \begin{equation*}
                q_\pm = k_\pm-p_\mp\,, \quad
                r_\pm = \frac{p_\pm-q_\mp}{2}\,,\quad
                r'_\pm = \frac{p_\pm-q_\pm}{2}\,.
            \end{equation*}
            For explicit calculations we work in the Breit frame~\eqref{breitframe} where ingoing and outgoing nucleon have opposite $3$-momenta
            and the photon consequentially carries zero energy.

    \subsection{Quark-photon vertex} \label{app:qpv}

            The general expression for the quark-photon vertex $\Gamma^\mu_\text{q}(k_+,k_-)$ is derived from the Ward-Takahashi identity
                  \begin{equation}\label{qpv-wti}
                      Q^\mu \,\Gamma^\mu_\text{q}(k_+,k_-) = S^{-1}(k_+)-S^{-1}(k_-)
                  \end{equation}
            and by imposing regularity at $Q^2=0$. It is given by
                  \begin{equation}\label{vertex:BC}
                      \Gamma^\mu_\text{q}(k_+,k_-) =   i\gamma^\mu\,\Sigma_A + 2 k^\mu (i\Slash{k}\, \Delta_A  + \Delta_B) + \widetilde{\Gamma}^{\mu}_T\,,
                  \end{equation}
            where $k_+$ and $k_-$ are outgoing and incoming quark momenta, $k=(k_+ + k_-)/2$, $Q=k_+ - k_-$, and
                  \begin{equation*}\label{QPV:sigma,delta}
                     \Sigma_F := \frac{F(k_+^2)+F(k_-^2)}{2} , \quad  \Delta_F := \frac{F(k_+^2)-F(k_-^2)}{k_+^2-k_-^2},
                  \end{equation*}
            with the quark propagator's dressing functions $A(p^2)$ and $B(p^2)=M(p^2)A(p^2)$.
            The first part is the Ball-Chiu vertex~\cite{Ball:1980ay}.
            The transverse contribution $\widetilde{\Gamma}^{\mu}$ is not constrained by the WTI except for the fact
            that it must vanish at $Q^2=0$. In the present work we use a phenomenological ansatz which is modeled after a RL-truncated
            inhomogeneous BSE solution for the quark-photon vertex~\cite{Maris:1999bh}.
            The latter includes a self-consistently generated vector-meson pole at $Q^2=-m_\rho^2$
            whose contribution significantly increases the charge radii of pseudoscalar and vector mesons~\cite{Maris:1999bh,Bhagwat:2006pu}.
            The ansatz reads
            \begin{equation}\label{vertex:simulate}
                \widetilde{\Gamma}^{\mu}_T =  - \frac{1}{g_\rho}\, \frac{x}{x+1} \, e^{-g(x) }\, T^{\mu\nu}_Q \,\Gamma^\mu_\text{vc},
            \end{equation}
            where $\Gamma^\mu_\text{vc}$ is the $\rho-$meson amplitude
            as obtained from its homogeneous Bethe-Salpeter equation, $x=Q^2/m_\rho^2$, and $g_\rho=\sqrt{2}\, m_\rho/f_\rho$ with the computed $\rho$ mass and decay constant.
            An additional function $e^{-g(x)}$ was implemented to optimize agreement with the vertex-BSE solution at low $Q^2$ and nucleon form factor phenomenology at intermediate $Q^2$,
            see \cite{Eichmann:2008ef} for details.

    \subsection{Seagulls}

            The seagull vertices $M^{\mu,\alpha}$ satisfy a Ward-Takahashi identity similar to Eq.\,\eqref{qpv-wti}
            which involves differences of axial-vector diquark amplitudes~\cite{Wang:1996zu,Oettel:1999gc,Eichmann:2007nn}.
            The resulting vertex as derived from this WTI has again the form
            \begin{equation}\label{ff:seagull}
                M^{\mu,\alpha} = M^{\mu,\alpha}_\text{WTI} + \widetilde{M}^{\mu,\alpha}_T \,.
            \end{equation}
            In a similar spirit as before we include a phenomenological transverse $\rho-$meson part $\widetilde{M}^{\mu,\alpha}_T$
            which optimizes agreement with nucleon form factors
            at intermediate $Q^2$ but is irrelevant on the domain $Q^2 \lesssim 1$ GeV$^2$.
            The detailed structure and derivation of \eqref{ff:seagull} can be found in App. A.8 of Ref.~\cite{Eichmann:2009zx}.

    \subsection{Diquark-photon vertex} \label{app:dqpv}

             The procedure which allows to derive the electromagnetic current for a composite particle can be
             generalized for an arbitrary composite vertex at off-shell momenta. Once the propagator of the composite object
             is known, the vertex is constructed as the 'gauged' inverse propagator,
             i.e. such that the photon couples to all its constituents.
             The Ward-Takahashi identity for the vertex is then automatically satisfied~\cite{Kvinikhidze:1999xp,Oettel:2002wf}.

             Starting from the expression for the axial-vector diquark propagator, Eq.\,\eqref{dqprop},
             the axial-vector diquark-photon vertex is constructed by coupling the photon to each
             of the constituents of the one- and two-loop integrals $F_{\mu\nu}$ and $Q_{\mu\nu}$.
             These involve the impulse-approximation coupling to the quark, specified by the quark-photon vertex, and the
             coupling to the diquark amplitudes, i.e. the seagulls. The seagull terms vanish on the diquark's mass shell
             --- hence the RL-consistent meson- or diquark-photon current is described by the impulse approximation ---
             but must be considered for off-shell momenta in order to satisfy the WTI.

             Having determined the quark-photon and seagull vertices previously, the diquark-photon vertex is
             completely specified. It is given in App. A.7 of Ref.\,\cite{Eichmann:2009zx}.

    \subsection{Color, flavor and charge coefficients}\label{app:color-flavor-charge}

             The current matrix diagrams of App.~\ref{app:currentdiagrams} still have to be equipped with color and flavor-charge coefficients.
             The full Dirac, color and flavor structure of the $\Delta$ quark-diquark amplitude is given by
                        \begin{equation}
                            \Phi^{\mu\nu}(p,P) \,\otimes\, \frac{\delta_{AB}}{\sqrt{3}} \,\otimes\, \mathsf{t^i_{ab}}\,,
                        \end{equation}
             where $\Phi^{\mu\nu}$ is the Dirac-momentum part of Eq.\,\eqref{deltaamplitdecompos} and $\delta_{AB}$ its color factor,
             with $A,B=1,2,3$ corresponding to the quark and diquark legs.
            The flavor matrices $\mathsf{t^i_{ab}}$, where the index $\mathsf{i}=1,2,3$ belongs to the axial-vector diquark's three symmetric isospin-1 states,
            the index $\mathsf{a}=1,2$ to the quark's two isospin-$\nicefrac{1}{2}$ states and the index $\mathsf{b}=1\dots 4$ to the four isospin-$\nicefrac{3}{2}$ states of the $\Delta$,
            are given by
 \renewcommand{\arraystretch}{1.0}
                \begin{equation*}
                      \begin{array}{l}
                         \mathsf{t^1} := \left( \begin{array}{cccc} 1  & 0 & 0 & 0 \\ 0 & \textstyle\frac{1}{\sqrt{3}} & 0 & 0 \end{array}\right), \\[0.5cm]
                         \mathsf{t^3} := \left( \begin{array}{cccc} 0  & 0 & \textstyle\frac{1}{\sqrt{3}} & 0 \\ 0 & 0 & 0 & 1 \end{array}\right),
                      \end{array} \quad
                      \mathsf{t^2} := \left( \begin{array}{cccc} 0  & \textstyle\sqrt{\frac{2}{3}} & 0 & 0 \\ 0 & 0 & \textstyle\sqrt{\frac{2}{3}} & 0 \end{array}\right).
                \end{equation*}
             Denoting the four isospin-$\nicefrac{3}{2}$ unit vectors $\mathsf{e^r}$ by
                \begin{equation*}
                      \mathsf{e^{++}} = \left( \begin{array}{c} 1 \\ 0 \\ 0 \\ 0 \end{array} \right)\!, \;
                      \mathsf{e^{+}} =  \left( \begin{array}{c} 0 \\ 1 \\ 0 \\ 0 \end{array} \right)\!, \;
                      \mathsf{e^{0}} =  \left( \begin{array}{c} 0 \\ 0 \\ 1 \\ 0 \end{array} \right)\!, \;
                      \mathsf{e^{-}} =  \left( \begin{array}{c} 0 \\ 0 \\ 0 \\ 1 \end{array} \right)\!,
                \end{equation*}
            the isospin-$\nicefrac{1}{2}$ basis by
                \begin{equation*}
                      \mathsf{u} = \left( \begin{array}{c} 1 \\ 0  \end{array} \right)\!, \quad
                      \mathsf{d} =  \left( \begin{array}{c} 0 \\ 1 \end{array} \right)\!,
                \end{equation*}
            and the isospin-triplet diquark matrices by
            \begin{equation*}\label{dq:flavormatrices}
                 \mathsf{s^1} = \mathsf{uu^\dag}\,, \quad
                 \mathsf{s^2} = \frac{1}{\sqrt{2}}\left(\mathsf{ud^\dag+du^\dag}\right), \quad
                 \mathsf{s^3} = \mathsf{dd^\dag}\,,
            \end{equation*}
            the contraction $\mathsf{t^i}\,\mathsf{e^r}$ represents the remainders upon removing the diquark flavor matrices $\mathsf{s^i}$
            from the full three-quark flavor wave function $\sum_{i=1}^3\mathsf{s^i} \otimes  \mathsf{t^i}\,\mathsf{e^r}$.
            The corresponding 'super'-vectors $\Delta^{++}, \Delta^{+}, \Delta^{0}, \Delta^{-}$
            whose entries are $(\Delta^\mathsf{r})^\mathsf{i} = \mathsf{t^i}\,\mathsf{e^r}$ are then given by
 \renewcommand{\arraystretch}{1.2}
                \begin{equation*}
                      \left( \begin{array}{c} \mathsf{u} \\ 0 \\ 0 \end{array} \right),  \quad
                      \left( \begin{array}{c} \textstyle\frac{1}{\sqrt{3}} \,\mathsf{d} \\ \textstyle\sqrt{\frac{2}{3}} \,\mathsf{u} \\ 0 \end{array}\right),  \quad
                      \left( \begin{array}{c} 0 \\ \textstyle\sqrt{\frac{2}{3}} \,\mathsf{d} \\ \textstyle\frac{1}{\sqrt{3}} \,\mathsf{u} \end{array}\right),  \quad
                      \left( \begin{array}{c} 0 \\ 0 \\ \mathsf{d} \end{array} \right).
                \end{equation*}

             These definitions allow to write down concise expressions for the color and flavor traces of the electromagnetic current.
             The color traces for the impulse approximation and exchange/seagull diagrams are given by
             \begin{equation*}
                 \frac{\delta_{BA}}{\sqrt{3}}\,\frac{\delta_{AB}}{\sqrt{3}}=1\;, \quad
                 \frac{\delta_{BA}}{\sqrt{3}} \, \frac{\varepsilon_{AED}}{\sqrt{2}}\,\frac{\varepsilon_{CEB}}{\sqrt{2}}\,\frac{\delta_{CD}}{\sqrt{3}}=-1\;,
             \end{equation*}
             respectively.
             The flavor-charge $4\times 4$ matrices for the quark-photon, diquark photon and exchange diagrams read:
             \begin{equation*} \label{flavorcharge:1}
                 \sum_{i=1}^3 \mathsf{t_i^\dag} \,\mathsf{Q}\, \mathsf{t_i}\,, \quad
                 \sum_{i,j=1}^3 \mathsf{t_i^\dag} \, \mathsf{t_j} \,2\, \text{Tr}\left\{ \mathsf{s_i^\dag}\,\mathsf{s_j}\,\mathsf{Q} \right\}, \quad
                 \sum_{i,j=1}^3 \mathsf{t_i^\dag} \,\mathsf{s_j}\,\mathsf{Q}\, \mathsf{s_i^\dag}\,\mathsf{t_j}\,,
             \end{equation*}
             where $\mathsf{Q}=\textit{diag}\,(q_u,q_d)$ is the quark charge matrix.
             Similar expressions arise for the seagulls, see App.~A.9 of Ref.~\cite{Eichmann:2009zx} for details.
             As expected from isospin symmetry, the flavor-charge matrices for all diagrams are proportional to $\textit{diag}\,(2,1,0,-1)$,
             with a coefficient $\nicefrac{2}{3}$ for the diquark part of the impulse approximation and $\nicefrac{1}{3}$ for all other diagrams.
             Consequently, the full Dirac-color-flavor-charge
             current for each member of the $\Delta$ multiplet with charge $e_\mathsf{r}$ is given by
             \begin{equation*}
                 J_\mathsf{r}  = \frac{e_\mathsf{r}}{3} \left( J^\text{IMP-Q}  + 2 J^\text{IMP-DQ} - J^\text{EX}  - J^\text{SG} - J^{\overline{\text{SG}}}  \right),
             \end{equation*}
             and the four different $\Delta$ states only differ by their charge.

\end{appendix}

%%%%%%%%%%%%%%%%%%%%%%%%%%%%%%%%%%%%%%%%%%%%%%%%%%%%%%%%%%%%%%%%%%%%%%%%%%%%%%%%%%%%%%%%%%%%%%%%%%%%%%%%%%%%%%%%%%%%%%%%%%%%%%%%%%%%%%%%%%%%%%%%%%%%%%%%%%%%%%%%%%%%%%%%%%%%%
%%%%%%%%%%%%%%%%%%%%%%%%%%%%%%%%%%%%%%%%%%%%%%%%%%%%%%%%%%%%%%%%%%%%%%%%%%%%%%%%%%%%%%%%%%%%%%%%%%%%%%%%%%%%%%%%%%%%%%%%%%%%%%%%%%%%%%%%%%%%%%%%%%%%%%%%%%%%%%%%%%%%%%%%%%%%%
%%%%%%%%%%%%%%%%%%%%%%%%%%%%%%%%%%%%%%%%%%%%%%%%%%%%%%%%%%%%%%%%%%%%%%%%%%%%%%%%%%%%%%%%%%%%%%%%%%%%%%%%%%%%%%%%%%%%%%%%%%%%%%%%%%%%%%%%%%%%%%%%%%%%%%%%%%%%%%%%%%%%%%%%%%%%%

\bigskip

\bibliographystyle{apsrev-mod}

\bibliography{lit-delta}

\end{document}